\documentclass{emulateapj}
\usepackage{macros}

\shorttitle{GC Magnetic Field Organization}
\shortauthors{Law et al.}

\begin{document}

\title{A Constraint on the Organization of the Galactic Center Magnetic Field Using Faraday Rotation}

\author{C. J. Law\altaffilmark{1}}
\author{M. A. Brentjens\altaffilmark{2}}
\author{G. Novak\altaffilmark{3}}
\altaffiltext{1}{Radio Astronomy Lab, University of California, Berkeley, CA, USA;  claw@astro.berkeley.edu}
\altaffiltext{2}{ASTRON, Dwingeloo, Netherlands}
\altaffiltext{3}{Department of Physics and Astronomy, Northwestern University, Evanston, IL, USA}

\begin{abstract}
We present new 6 and 20 cm Very Large Array (VLA) observations of polarized continuum emission of roughly 0.5 square degrees of the Galactic center (GC) region.  The 6 cm observations detect diffuse linearly-polarized emission throughout the region with a brightness of roughly 1 mJy per 15\arcsec$\times$10\arcsec\ beam.  The Faraday rotation measure (RM) toward this polarized emission has structure on degree size scales and ranges from roughly +330 rad m$^{-2}$ east of the dynamical center (Sgr A) to --880 rad m$^{-2}$ west of the dynamical center.  This RM structure is also seen toward several nonthermal radio filaments, which implies that they have a similar magnetic field orientation and constrains models for their origin.  Modeling shows that the RM and its change with Galactic longitude are best explained by the high electron density and strong magnetic field of the GC region.  Considering the emissivity of the GC plasma shows that while the absolute RM values are indirect measures of the GC magnetic field, the RM longitude structure directly traces the magnetic field in the central kiloparsec of the Galaxy.  Combining this result with previous work reveals a larger RM structure covering the central $\sim2$\sdeg\ of the Galaxy.  This RM structure is similar to that proposed by Novak and coworkers, but is shifted roughly 50 pc west of the dynamical center of the Galaxy.  If this RM structure originates in the GC region, it shows that the GC magnetic field is organized on $\sim300$ pc size scales.  The pattern is consistent with a predominantly poloidal field geometry, pointing from south to north, that is perturbed by the motion of gas in the Galactic disk.
\end{abstract}

\section{Introduction}
Diverse observations of the center of the Milky Way have found evidence for magnetic field strengths from 10--1000 $\mu$G with both poloidal and planar geometries.  Zeeman splitting of H I absorption lines shows that mG-strength magnetic fields exist in the central 2 pc \citep{p95}.  On the basis of their submillimeter polarimetric maps of the Galactic center region, \citet{c03} argue that molecular clouds in the central 100 pc have mG-strength fields oriented parallel to the plane of the Galaxy.  At the same time, the detection of diffuse, polarized radio continuum emission implies that the central few hundred parsecs is permeated by magnetic field of strength 10 to 100 $\mu$G \citep{h92}.

The most striking polarized structures in the GC region were discovered in radio continuum images:  the nonthermal radio filaments \citep[NRFs;][]{y84}.  NRFs are long (several parsecs), polarized filaments found only in the central few degrees of our Galaxy and believed to be physically in the central few hundred parsecs \citep{l01,n04,y04,l89}.  Their tendency to align perpendicular to the Galactic plane shows that the GC region has some poloidal (i.e., vertical) magnetic field component with a local strength 1 mG \citep{y97}.  This vertical structure is interesting in light of infrared polarimetry showing that the magnetic field tends to have a toroidal configuration in the plane, but becoming poloidal at elevations above $0\ddeg4$\ \citep{n10}.

Despite the wealth of observations, it has been difficult to merge these observations into a coherent model for the Galactic center (GC) magnetic field.  How are the magnetic fields measured globally and locally related to each other \citep{la05,f09,c10}?  What physical processes create the NRFs and determine their orientations \citep{l06}?  Is the current state of the GC normal or does it represent a short phase of its evolution \citep{m96,gcl_all}?  Furthermore, the complexity of the range of polarimetric observations (some measure line-of sight field, some measure total field) argues for simulations to aid interpretation.

To understand the structure and strength of the GC magnetic field, we present new observations and modeling of the polarized continuum emission toward the GC region with the VLA.  The observations were originally conducted in a study of the GC Lobe, a degree-tall, loop-like structure spanning the central degree of the GC region \citep{s84, b03, gcl_all}.  In \S\ \ref{poln_obs}, the observations are described; \S\ \ref{poln_analysis} discusses some of the techniques used to analyze the polarized emission.  This survey is the largest-area, interferometric survey of diffuse polarized emission ever done in the GC region.  Section \ref{poln_results} describes the detection of extended, polarized emission throughout the region and the large-scale rotation measure (RM) structure seen towards it.  Section \ref{poln_discussion} uses the observed RM to constrain a simple model of the Galaxy's electron density and magnetic field.  Modeling of the emission and Faraday rotation argues that the GC magnetic field geometry is predominantly poloidal with a perturbation by motion of gas in the disk of the Galaxy.

\section{Observations and Data Reduction}
\label{poln_obs}
Between January and August of 2004, we surveyed the GC region with the VLA at 6 cm in the DnC configuration and at 20 cm in the CnB, DnC, and D configurations.  The goal of the observation was to create wide-field mosaics of the GC lobe, as described in \citet{gcl_vla}.  That paper presents catalogs of discrete polarized and unpolarized sources in the survey;  extended polarized emission, particularly at 6 cm, is discussed here.  The 6 cm observations covered roughly half a square degree from $l=$359\ddeg2 to 0\ddeg2, $b=$0\ddeg2 to 0\ddeg7.  Critically for the present work, the default continuum mode observed with two, adjacent 50 MHz bands centered at 4.835 and 4.885 GHz.

Observations of J1751--253 were used for phase calibration, while J1331+305 (3C 286) was used for flux calibration.  Observations of the unpolarized phase calibrator covered a parallactic angle range of 80\sdeg.  This is wide enough to measure receiver ``leakage'', the detection of left-circular polarization by the right-circular receiver and vice versa \citep{c99}.  After applying the leakage corrections to the scan of 1331+305, the phase delay between left and right polarizations was set to produce the known polarization angle of 66\sdeg.\footnote{See \url{http://www.vla.nrao.edu/astro/calib/manual/polcal.html}; R. Perley \& N. Killeen, private communication}

Images were produced with AIPS \footnote{See \url{http://www.aips.nrao.edu}} using both the multi-resolution and the standard CLEAN algorithms.  The resulting mosaics and derived properties were similar within their errors.  The final mosaics presented here were deconvolved with a multi-resolution CLEAN algorithm.  The Stokes Q and U images were cleaned independently with resolutions of 1, 3, and 9 times the beam size to produce a single image per Stokes parameter.  The entire primary beam was cleaned until the maximum residual brightness was less than the noise level outside the primary beam.  The same number of iterations was used to clean both bands.  Images were restored with a single beam of size 15\arcsec$\times$10\arcsec\ with PA$=70$\sdeg, which is representative of the whole mosaic.

\begin{figure}[tbp]
\includegraphics[width=0.5\textwidth, trim=0 150 0 100, clip]{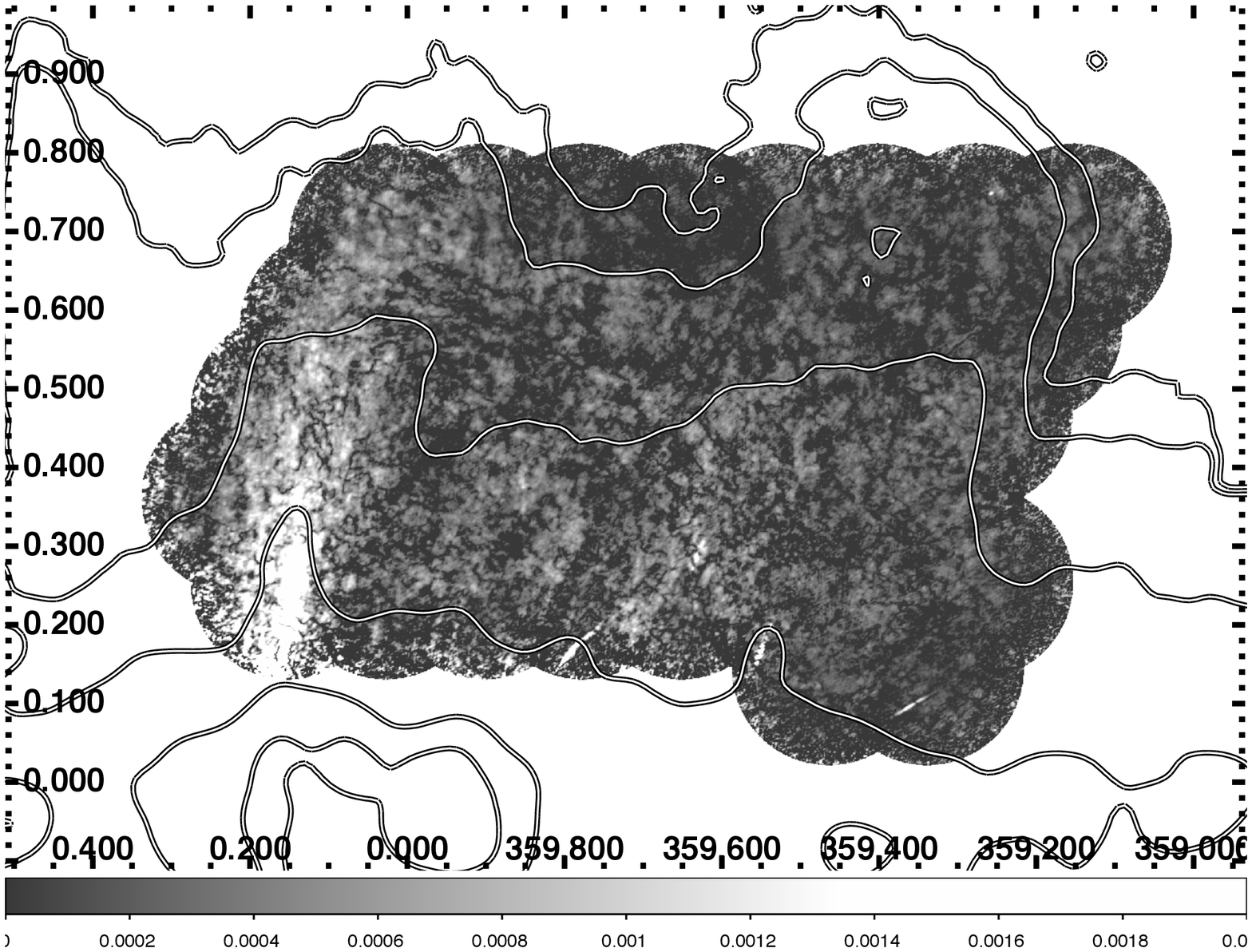}
\caption{Mosaic of 6 cm polarized intensity observed in 42 VLA pointings with contours of 6 cm total intensity from the GBT with Galactic coordinates \citep{gcsurvey_gbt}.  The beam size is 15\arcsec$\times$10\arcsec\ with a position angle of 70\sdeg.  Gray scale shows VLA 6 cm polarized intensity from 0 to 2 mJy beam$^{-1}$ as indicated on the colorbar in units of Jy.  Contours show GBT 6 cm brighness at $33*3^n$ mJy per 150\arcsec\ circular beam, for $n=0-4$.  Most of the polarized emission seen is significant. \label{poln_polc}}
\end{figure}

\begin{figure}[tbp]
\begin{center}
\includegraphics[height=0.4\textwidth,angle=270]{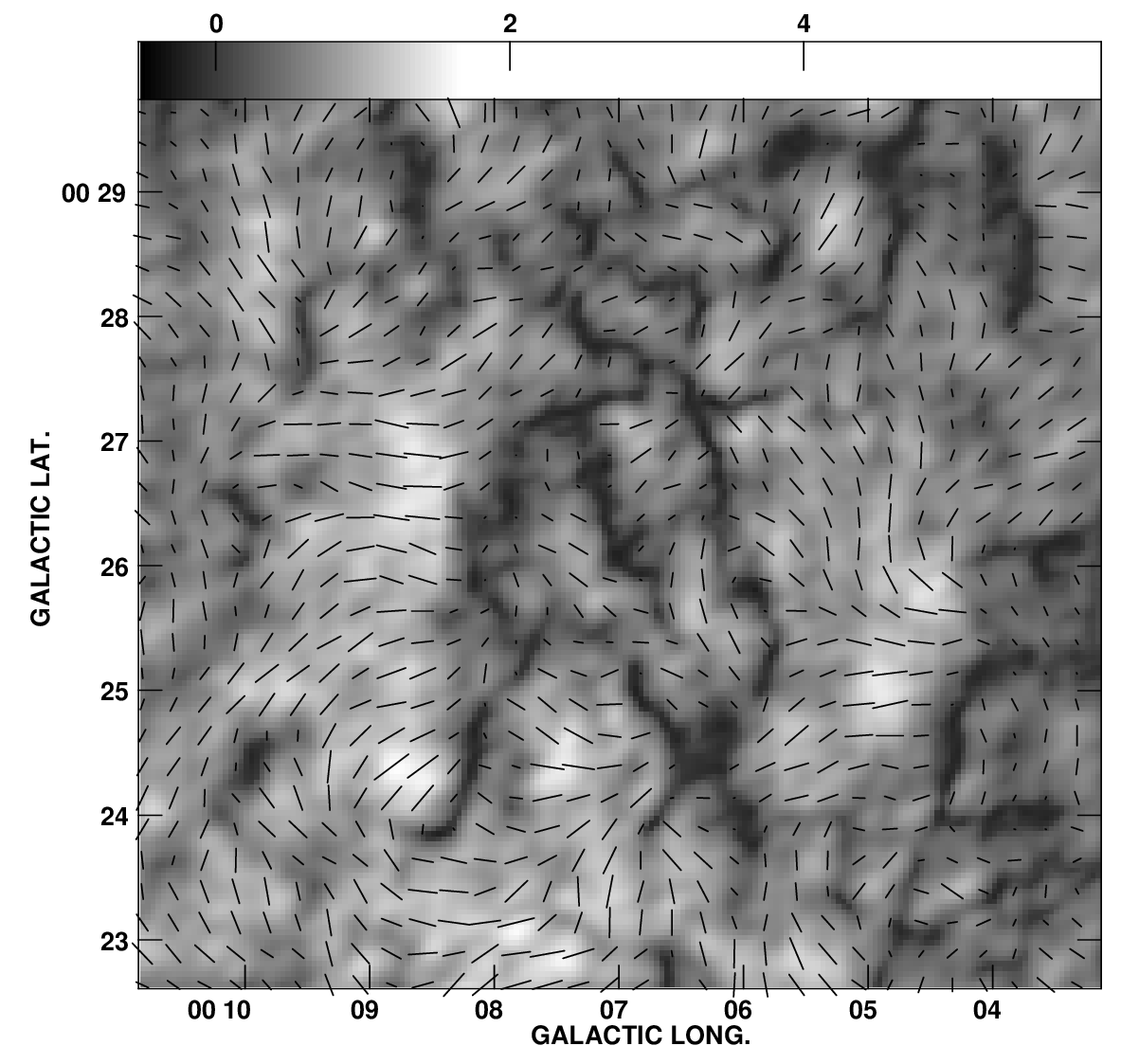}
\end{center}
\caption{Map of the 6 cm polarized intensity near the eastern edge of the survey, where depolarized ``canals'' are evident.  The lines show the polarization angle of the emission at 4.835 GHz.  The polarization angle changes by 90\sdeg\ across the canals, indicating depolarization within the beam by small-scale changes in the Faraday-rotating medium \citep{h04}. \label{canal}}
\end{figure}

The Stokes Q and U images were subsequently primary-beam corrected and combined to form mosaics for each band.  Figure \ref{poln_polc} shows the polarized intensity mosaic after averaging over both bands.  The polarized intensity is visible on scales of a few arcminutes because it is laced with depolarized ``canals'' \citep{w93,y86,h04}.  Figure \ref{canal} shows an example of canals in the polarized emission in the eastern half of the survey.

To correct for noise bias in maps of polarized emission, a noise mosaic was constructed by quadratically adding noise images from each field.  A noise image was created for each field by applying the primary beam correction to an image with each pixel value set to the image noise.  The noise level for each field was measured outside the primary beam, since the image centers are filled with emission.  Observed noise values range from 50 to 120 $\mu$Jy.  The observed noise is a factor of 2--4 times higher than the theoretical sensitivity, which is consistent with the expected additional noise from sidelobes and calibration errors.  Finally, mosaics of polarized intensity, position angle, and their associated errors were created for each band.

Leakage calibration is most valid at the phase center, as errors are known to increase away from there.  The VLA position-dependent leakages are based in the antenna and induce a false linear polarization that is radially oriented \citep{c94,c99}.  The magnitude of the false polarization is roughly 3\% of Stokes I at the full-width at half-max (FWHM) of the primary beam at 1.4 GHz.  It is not measured at other frequencies, but we use 3\% as a rough estimate of errors at 5 GHz.  Since this error is antenna based, our observations covering a parallactic angle range of 80\sdeg\ reduces the effect by about 30\%.  More importantly, the diffuse polarized emission discussed in this work has no total intensity counterpart, so errors in Stokes Q and U scale with Stokes Q and U, instead of Stokes I \citep{s96}.  Furthermore, measurements of RM are not biased by the leakage, but the change in leakage with frequency, which tends to be smaller.  Considering all these effects, we expect position-dependent leakage errors to be less 3\% of Stokes Q and U, less than the typical flux calibration errors and unlikely to affect the results presented here.  Section \ref{poln_comparison} compares our results to previous GC polarimetry observations and generally confirms this assumption.

Finally, it is important to consider the fact that interferometric observations are not sensitive to emission on large angular scales \citep{h04,s09}.  Missing Q and U flux can create spurious polarization and bias RM values.  \citet{h04} show that a wide distribution of RM randomizes any uniform polarized background and reduces the missing flux.  The RM distribution observed here (described in detail in \S\ \ref{padianalysis}) has a width of about 500 rad m$^{-2}$ on size scales used in this study ($>$100\arcsec), which limits the missing flux to less than 0.2\%.  As an alternative derivation of missing flux, \citet{s09} show that a gradient in RM can shift the spatial scale at which polarized emission is visible.  For the RM gradient seen here ($\approx5$ rad m$^{-2}$ arcsec$^{-1}$), that technique predicts a shift in spatial scales from zero to about 1200 $\lambda$, larger than our shortest baseline.  Both techniques indicate that an insignificant amount of polarized flux is missed by the present observations.

\section{Analysis}
\label{poln_analysis}

\subsection{Polarization Angle Difference Across Bands}
\label{padianalysis}
\begin{figure}[tbp]
\begin{center}
\includegraphics[width=0.5\textwidth, trim=0 200 0 180, clip]{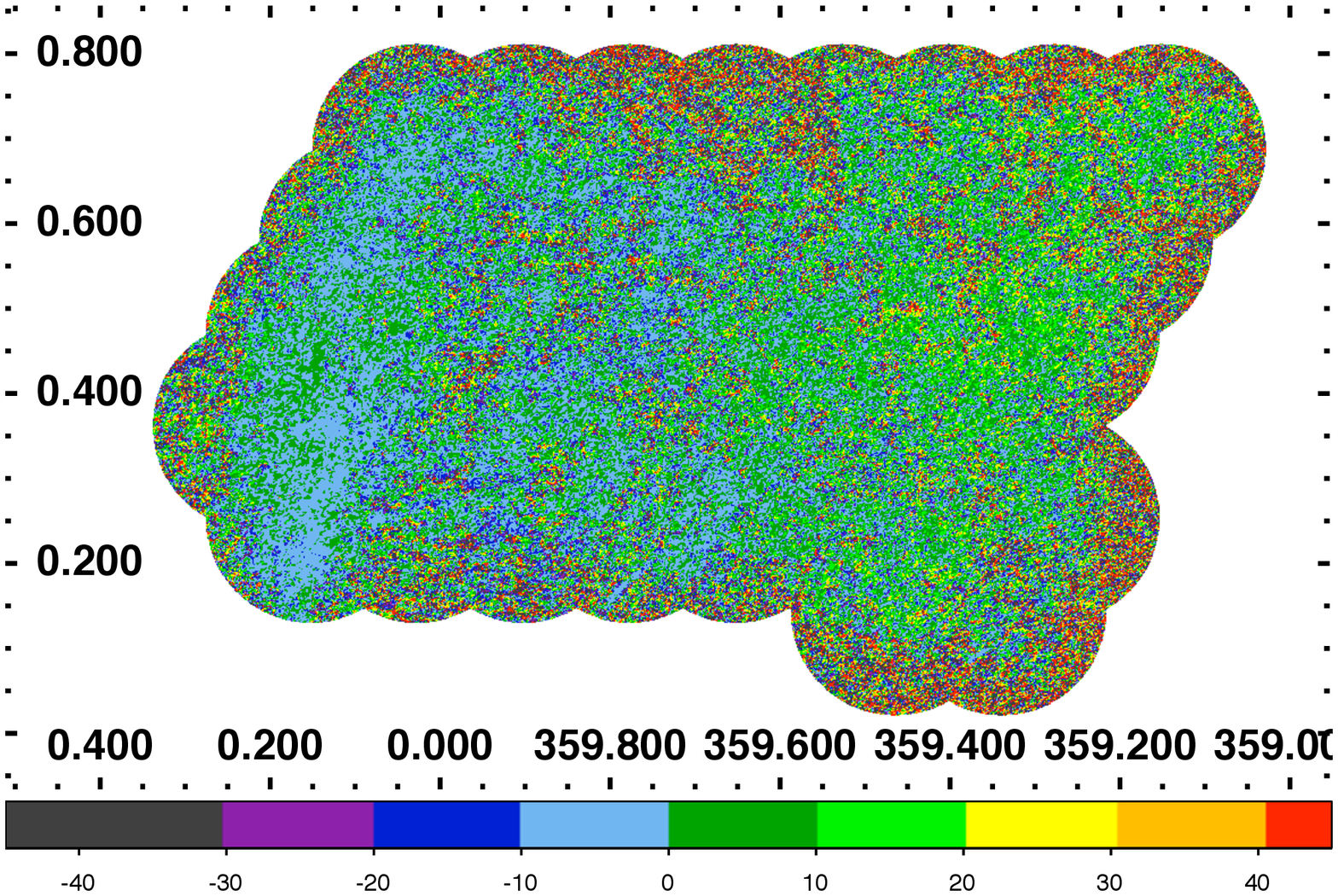}

\includegraphics[width=0.5\textwidth, trim=0 200 0 180, clip]{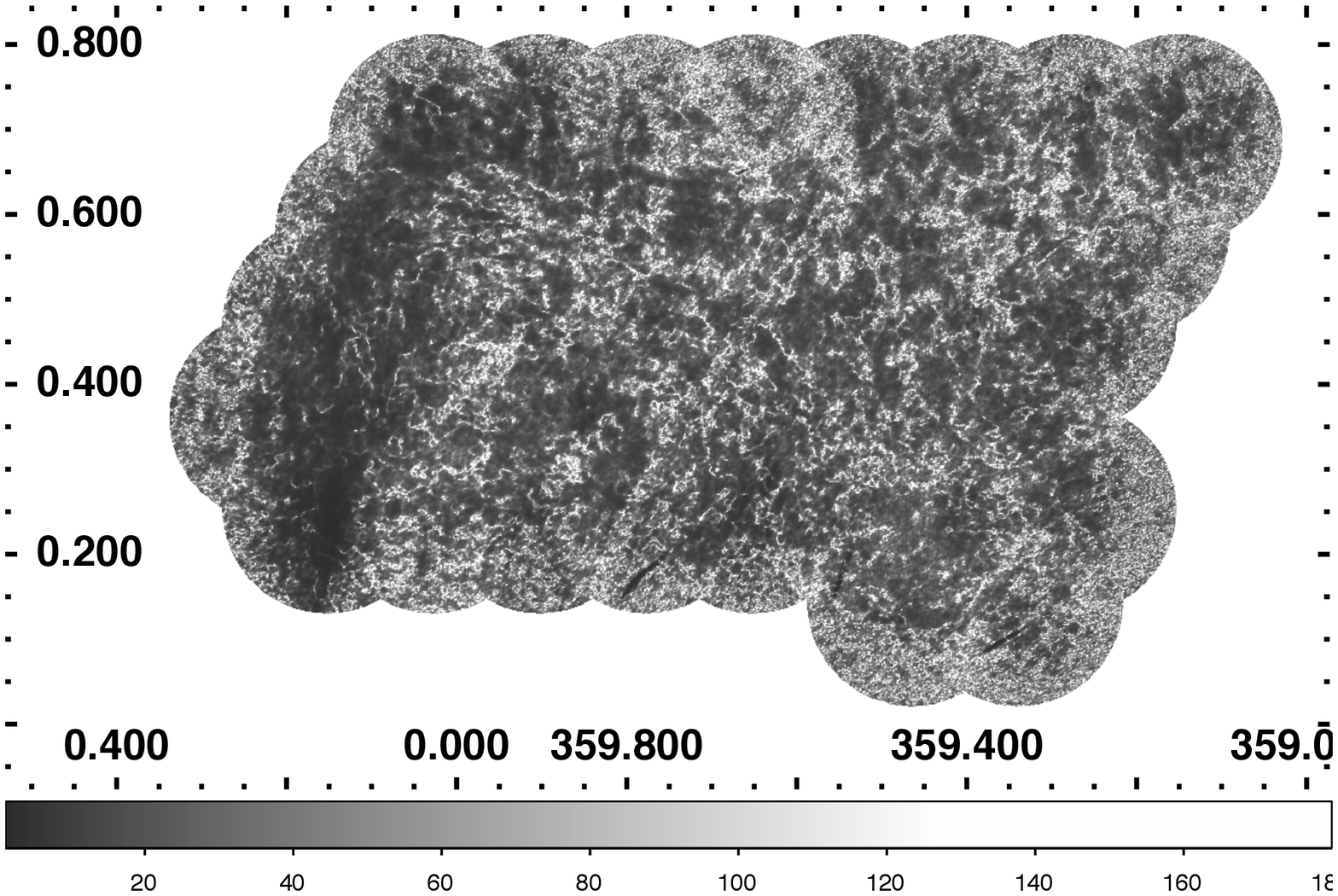}
\end{center}
\caption{\emph{Top}: Map of the polarization angle difference, or \dt, for the 6 cm survey of the GCL.  The color scale is linear according to the colorbar at the bottom in units of degrees;  1\sdeg is equivalent to a RM of --220 rad m$^{-2}$.  The edges of the map have large values of \dt\ because the sensitivity is low there.  \emph{Bottom}:  Map of the error in \dt.  The greyscale is linear, with values ranging from 1\ddeg2 to 180\sdeg.  The typical error in the \dt\ value is about 5\sdeg. \label{poln_padi}}
\end{figure}

To study the RM across this field, mosaics of the polarization angle were differenced between the two bands.  The polarization angle difference image (hereafter \dt\ image) was created by differencing the polarization angle images ($\theta_{\rm{4.885 GHz}}-\theta_{\rm{4.835 GHz}}$) and remapping each value of \dt\ to the range --90\sdeg\ to 90\sdeg.  Figure \ref{poln_padi} shows the \dt\ image and its error.  Observationally, the rotation measure is defined as RM$ = \Delta\theta/\Delta(\lambda^2)$.  Assuming this $\lambda^2$ law, a position angle difference of 1\sdeg\ corresponds to a rotation measure of --220 rad m$^{-2}$.

More generally, the observed polarization is the sum of polarized emission emitted with a range of RM \citep{b66,br05}.  Such complex sources can have non-quadratic changes in the polarization angle that can confuse a simple analysis.  In these situations, the pertinent physical quantity is the ``Faraday depth'':
\begin{equation}
\phi = 0.81 \int_{there}^{here} \! n_e \, \vec{B} \, \cdot d\vec{l} \, \rm{rad} \, \rm{m}^{-2},
\label{phieqn}
\end{equation}
\noindent where $n_e$ is in cm$^{-3}$, $\vec{B}$ is in G, and $d\vec{l}$ is in pc.  For simple physical distributions of $n_e$\ and $\vec{B}$, $\phi$\ is equal to RM.  However, robustly tying the RM to physical conditions in complex cases requires measurements at many wavelengths \citep{br05}.  Since we only have two wavelengths to study the polarization in this region, we instead use this formalism to define the limits of deriving physical conditions from the observed RM.

First, the formalism of \citet{br05} shows that the spacing of the bands in wavelength determines the ``RM resolution'' and possible $n\pi$ ambiguities.  For the two bands used here, the RM resolution is $4\times10^4$\ rad m$^{-2}$ and any aliasing occurs at RM$=n * 4\times10^4$ rad m$^{-2}$, for an integer $n$.  The RM expected in the GC region covered by this survey is typically $<2000$\ rad m$^{-2}$ \citep{y84,t86,r05}, so there is little chance of an $n\pi$ ambiguity.  Second, the bandwidth determines the amount of Faraday rotation within a band, which limits the maximum Faraday depth detectable to $\phi<2\times10^4$\ rad m$^{-2}$.  Finally, a source that emits over a range of Faraday depths, known as ``Faraday thick'', can be internally depolarized.  The maximum Faraday thickness detectable to the present observations is 830 rad m$^{-2}$.  Some sources, such as the Radio Arc \citep{y87,y88}, have an RM that changes by more than 830 rad m$^{-2}$, so parts of the GC region may be Faraday thick to our observations.

\begin{figure}[tbp]
\begin{center}
\includegraphics[scale=0.4]{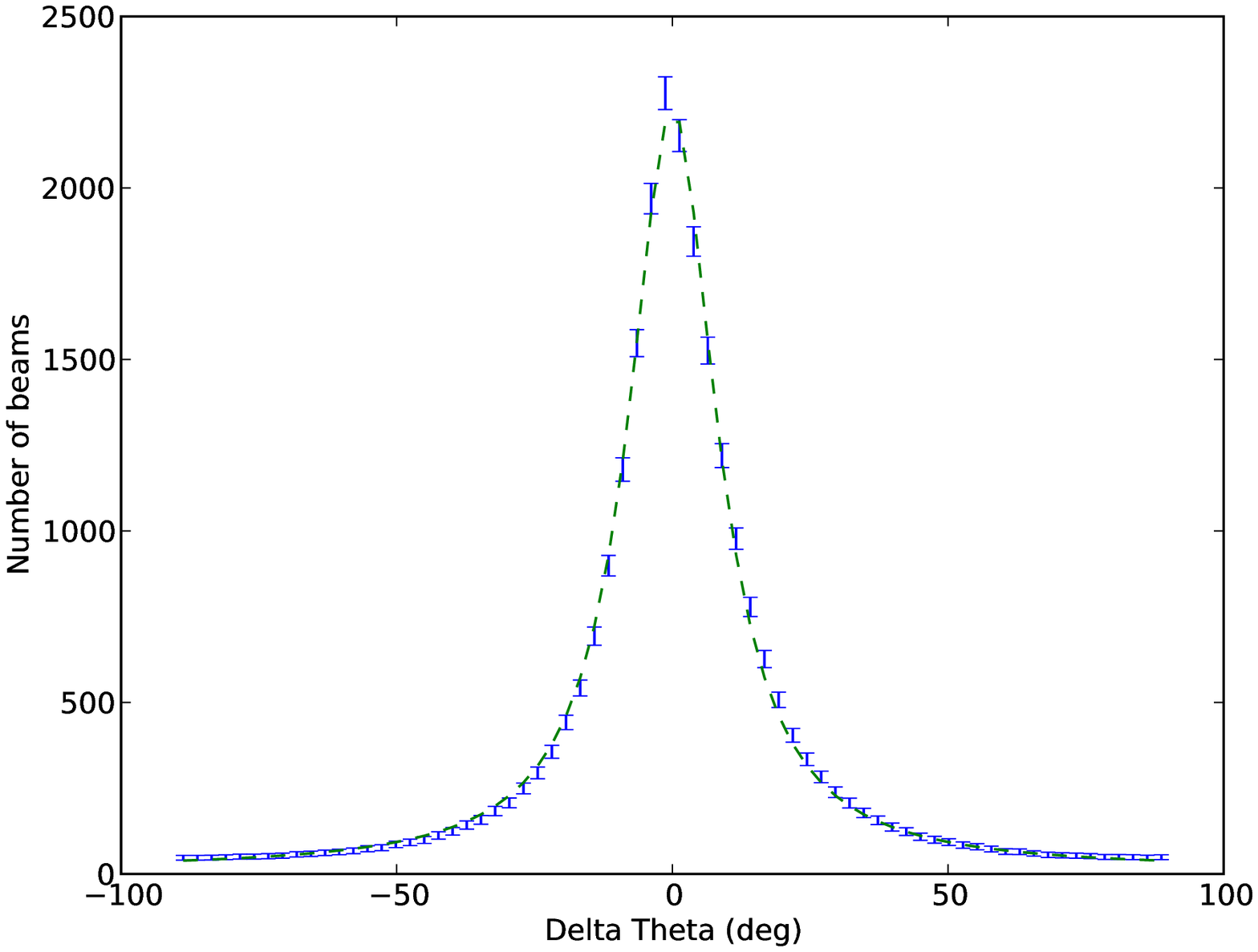}
\hfil
\includegraphics[scale=0.4]{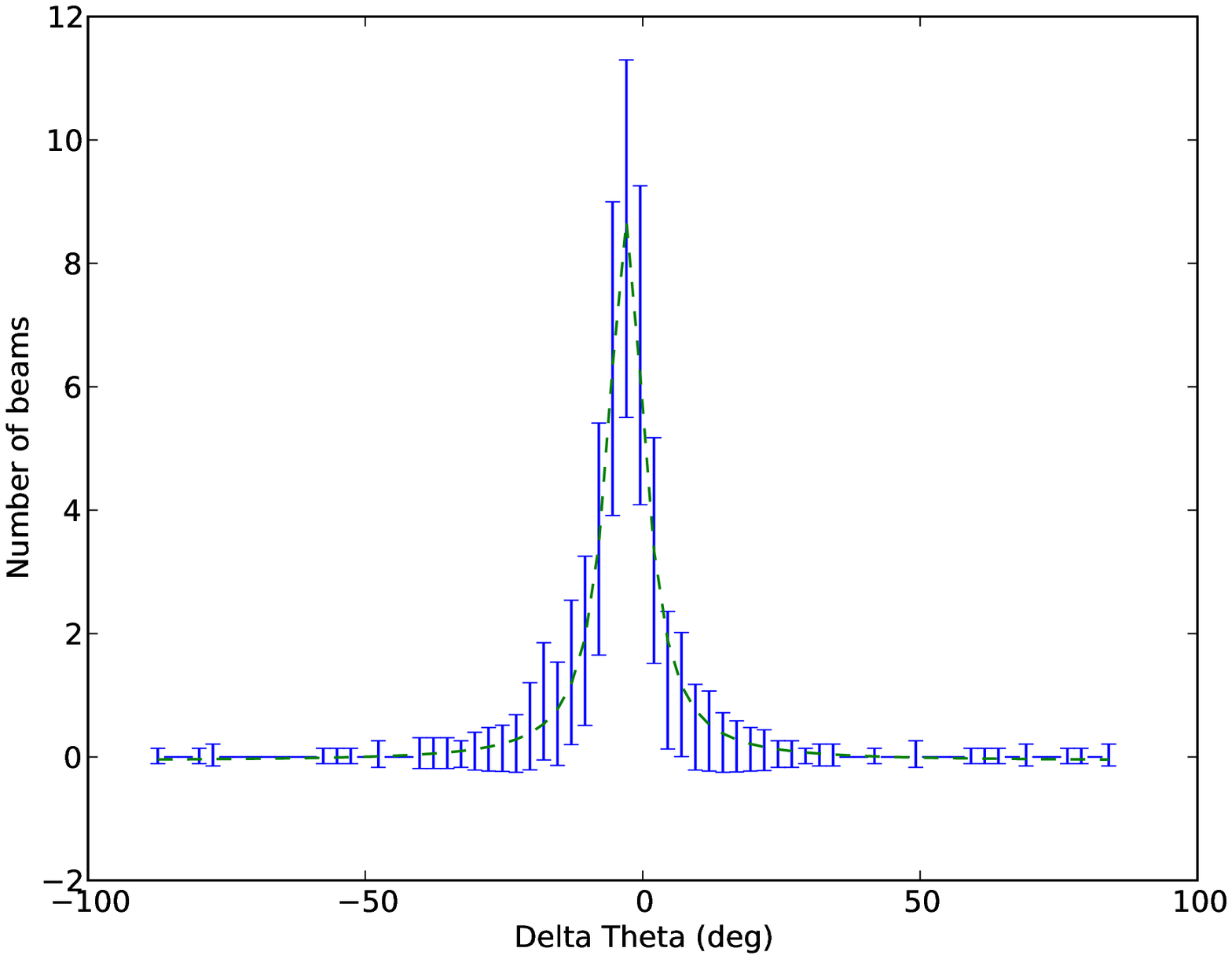}
\end{center}
\caption{\emph{Top}: Histogram of \dt\ in each independent beam in the survey region. The blue error bars show the \dt\ values assuming they follow a Poisson distribution and uses a bin size of 2\ddeg5 of \dt. The green dashed line shows the best-fit Lorentzian model, which has a peak of 2200 beams per degree, a half-width of about 10\sdeg, and a constant background of about 13 beams per degree of \dt.  \emph{Bottom}: A similar plot as shown at top, but for a 125\arcsec$\times$125\arcsec\ box in the eastern half of the survey.  This histogram shows the amount of data in each pixel of the smoothed maps shown below. \label{histcomp}}
\end{figure}

Figure \ref{histcomp} shows histograms of \dt\ (number of independent spatial beams per degree of \dt) from the entire survey and a smaller region.  Intrinsically, we expect the \dt\ histogram to have contributions from many distinct regions of varying peak \dt\ and width.  We found that a Lorentzian profile fits these heterogeneous distributions better than a single Gaussian.  For smaller regions, where \dt\ has a single-valued, noise-like distribution, the Lorentzian can also approximate a single Gaussian.  In the limit of a single-valued, noise-like \dt\ distribution, the Gaussian noise is equivalent to a Poisson distribution in the large-\emph{N} limit.  We use this similarity to approximate the \dt\ bin count errors as $\sigma_N = 1 + \sqrt{N + 0.75}$, where \emph{N} is the number of independent beams in a bin \citep{g86}. In \S \ref{poln_comparison}, we show that comparing RM measured histogram methods to previous work shows that the errors are conservative.

Aside from theoretical expectations, the distribution of \dt\ values shows that the apparent \dt\ can generally be reliably converted to RM.  Most values of \dt\ measured have small offsets from 0\sdeg, with $\sim$50\% within $\pm$10\sdeg, $\sim$75\% within $\pm$20\sdeg, and $\sim$90\% within $\pm$45\sdeg.  This is consistent with the $\sim$9\sdeg\ rotation expected for RM$\approx\pm2000$ rad m$^{-2}$.  The typical angle change is $\ll1$\ rad, so relative angles may be treated as roughly linearly distributed.  The best constraint on the mean \dt\ has a typical error of about 1\sdeg, or $220$ rad m$^{-2}$.  For a change of 1\sdeg\ between our two bands, the Faraday rotation from $\lambda=0$\ cm is 49\sdeg.  This typical uncertainty in \dt\ makes the calculation of the intrinsic polarization angle highly uncertain, so no such results are presented here.

Visualizing the images of \dt\ and RM is difficult, since the per-pixel sensitivity is poor and varies across the field of view.  Convolution and other image processing techniques can be used to extract this information even in poorly-calibrated VLA data \citep{r09}.  We tested two statistical techniques to spatially smooth the RM:  averaging and histogram fitting.  These methods and a comparison of their results are described in Appendix \ref{meandt}.  In general, the two methods have similar results.  The histogram-fitting method is less sensitive to outliers and has more conservative errors, so it is used in all results described below.

\subsection{Comparison to Earlier Work}
\label{poln_comparison}
Since this work is applying a relatively new technique to a complex region, it is important to test the results against known sources.  This section compares our results shown in Figures \ref{poln_polc} and \ref{poln_padi} to the RM for specific regions studied previously \citep{l01,t86,h92,r05,y97}.  \citet{l01} present images of 6 cm polarized intensity near the nonthermal radio filament G359.85+0.39 from VLA data with similar sensitivity and resolution as the present study.  The two surveys have similar brightness distributions and structure in the polarized emission, particularly the depolarized regions on the southeast and northeast sides of G359.85+0.39 \citep[see also][]{gcl_vla}.  The similarity shows that the calibration and imaging quality is similar to that of \citet{l01}.

\begin{figure}[tbp]
\begin{center}
\includegraphics[width=0.15\textwidth]{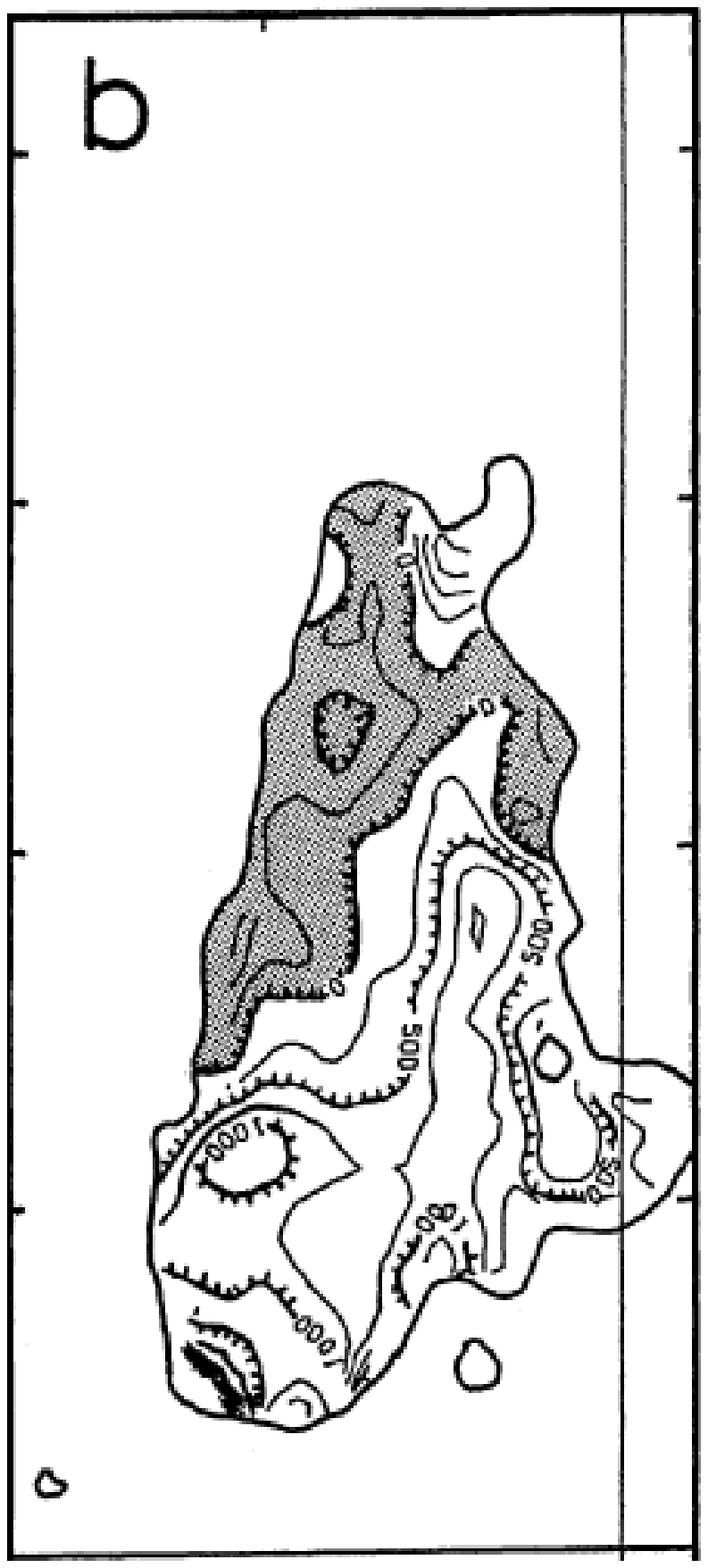}
\hfil
\includegraphics[width=0.28\textwidth, trim=25 50 0 0, clip]{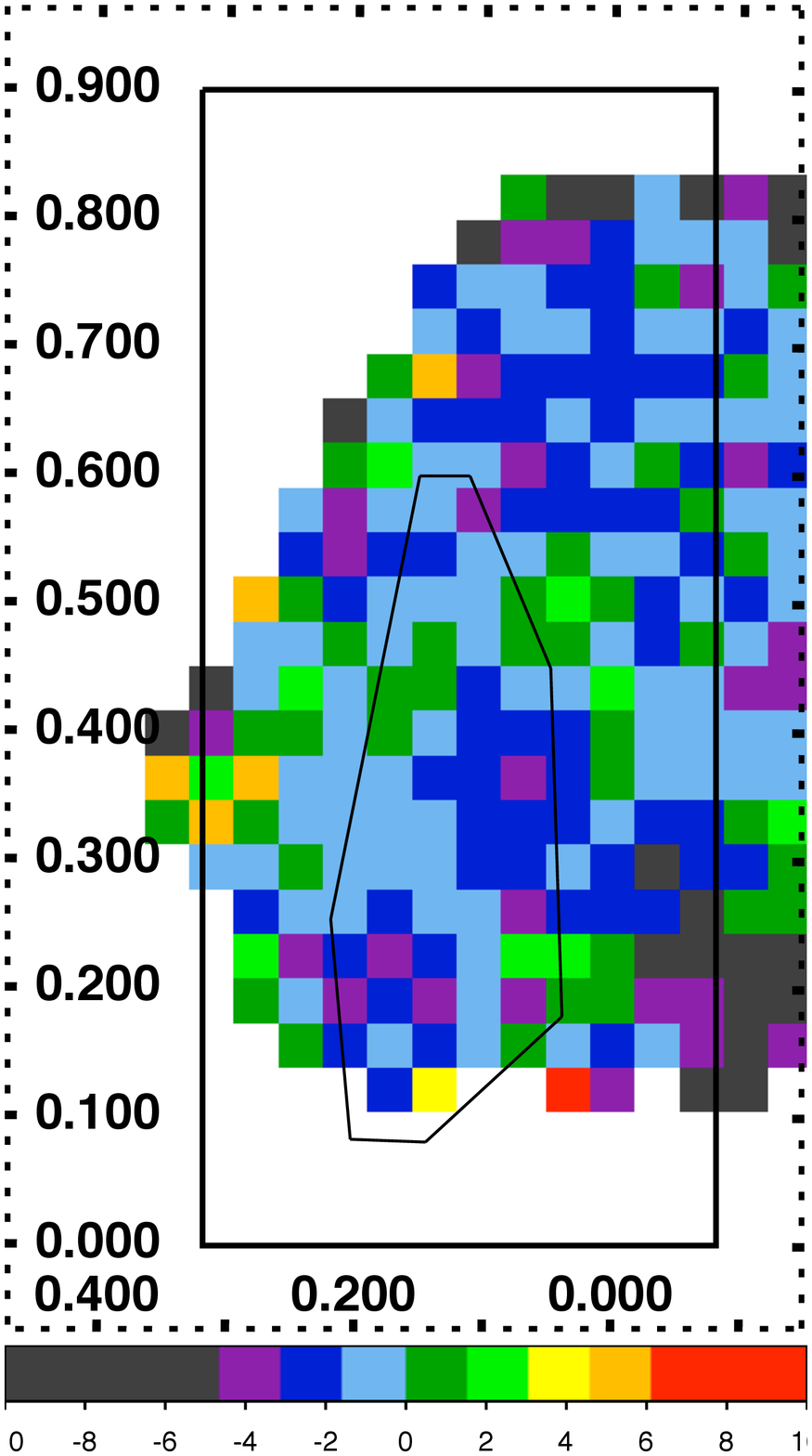}
\end{center}
\caption{\emph{Left}: Reprint of Fig. 4 from \citet{t86} showing the RM for the northern extension of the Radio Arc, as observed by the Nobeyama Radio Observatory near 3 cm with a beam size of 2\damin7.  The contours show steps of 250 rad m$^{-2}$, with the shaded region showing negative RM and unshaded region showing positive RM. \emph{Right}: \dt\ image from the present 6 cm, VLA survey smoothed by the histogram-fitting method with 125\arcsec$\times$125\arcsec\ boxes.  The colorbar shows the \dt\ value in units of degrees.  The rectangle shows the extent of the left figure on this figure and the irregularly-shaped box schematically shows the outermost contour from the left figure. \label{rmcomp}}
\end{figure}

\citet{h92} and \citet{t86} conducted independent, single-dish surveys near 3 cm, covering a few square degrees of the GC region.  Although depolarization is weaker near 3 cm and their beam is larger, there is general agreement between our Figure \ref{poln_polc} and their polarized intensity maps.  Figure \ref{rmcomp} shows a comparison of our RMs with the four-band measurements of \citet{t86}.  Near $(0\ddeg17,0\ddeg22)$\ and $(0\ddeg1,0\ddeg35)$, \citet{t86} find the RM has a maximum of +1000 rad m$^{-2}$, while the present survey finds a maximum of $770\pm110$ rad m$^{-2}$.  The maps are similar moving north across $(0\ddeg15,0\ddeg4)$, where the RM switches from positive to negative values;  \citet{t86} measure RM$\approx-250$ rad m$^{-2}$ while the present survey finds $-220\pm130$ rad m$^{-2}$.  There is some agreement at the northwestern edge of the polarized emission of the Radio Arc, shown in Figure \ref{rmcomp}, where the RM switches back to positive values.  The exact location of this second RM sign change is slightly different and may reflect the different RM depths each survey is sensitive to.

\citet{y97} presented a detailed study of the polarization properties of the nonthermal filament G359.54+0.18 (RF-C3) at 6 and 3.6 cm.  Figure 3 of that work has a similar 6 cm brightness and RM distribution as the present work, both presented here and in \citet{gcl_vla}.  The RM map of the filament shows three distinct, bright clumps each having relatively uniform values.  The morphology seen in the present survey is similar to that of \citet{y97}, although it had roughly three times better resolution (4\arcsec\ compared to 12\arcsec\ in the present work).  The first clump, at RA, Dec (B1950) = (17:40:41, --29:12:30) has RM$\approx-2700$ rad m$^{-2}$, compared to $-3960\pm1100$ rad m$^{-2}$ in the present survey.  The second clump, at (17:40:43, --29:12:40), has RM$\approx-2000$ rad m$^{-2}$, compared to $-2200\pm440$ rad m$^{-2}$ in the present survey.  The third clump, at (17:40:44,--29:12:45), has RM$\approx-1500$ rad m$^{-2}$, compared to $-1540\pm660$ rad m$^{-2}$ in the present survey.  We conclude that, in general, there is good agreement between the RM of the present survey and that of \citet{y97}.

In summary, the polarized intensity and RM of the present 6 cm survey shows good agreement with those of other surveys.  This is consistent with the fact the polarimetric leakgage is expected to have relatively little frequency structure for the VLA feed design \citep{c94,c99};  any systematic errors in the polarization angle are subtracted when forming the \dt\ image.  It also shows that histogram fitting of the \dt\ values is a reasonable estimate of the RM and its uncertainty at 6 cm in this region.

\section{Results}
\label{poln_results}

\subsection{Extended Polarized Emission}
The 6 cm polarized continuum intensity of the northern extension of the Radio Arc is several mJy beam$^{-1}$ and spans the entire eastern edge of the survey up to a latitude of $b\sim0\ddeg8$.  To test for frequency structure in the polarized intensity, the polarized intensity maps in the two bands were differenced.  The lack of diffuse emission in the difference map shows that the two maps have similar diffuse emission within roughly 1 mJy.  The comparable 20 cm mosaic of polarized continuum shows no extended emission down to a level of about 0.1 mJy beam$^{-1}$ \citep[more detail in][]{gcl_vla}.

For latitudes up to $b=0\ddeg3$, the polarized continuum emission seen in the 6 cm interferometric maps (Fig. \ref{poln_polc}) has a total intensity counterpart in the same data.  However, north of $b=0\ddeg3$, the total intensity counterpart is too extended to be detected by the VLA 6 cm observations.  Since the polarized emission is broken into small spatial scales (as shown in Figure \ref{canal}), it is detected throughout the region and the apparent polarization fraction often exceeds 100\%.  

To estimate the polarization fraction without the effect of missing flux, we compare the VLA polarized-intensity maps to continuum maps from the Green Bank Telescope \citep{gcsurvey_gbt}.  We convolve the VLA maps to the GBT resolution to estimate the polarization fraction;  this will be a lower limit, since the VLA emission is laced with depolarized canals.  At 6 cm, the peak polarization fraction is 25\% in the eastern half of the survey and 10\% in the western half of the survey.  These values are consistent with other single-dish surveys \citep{t86,h92}, which confirms the validity of techniques and maps of the VLA survey.  At 20 cm, the upper limit on the polarization fraction is roughly 1\% of the total intensity measured by the GBT.

\subsection{Degree-scale RM Structure}
\begin{figure}[tbp]
\begin{center}
\includegraphics[width=0.45\textwidth, trim=0 150 0 150, clip]{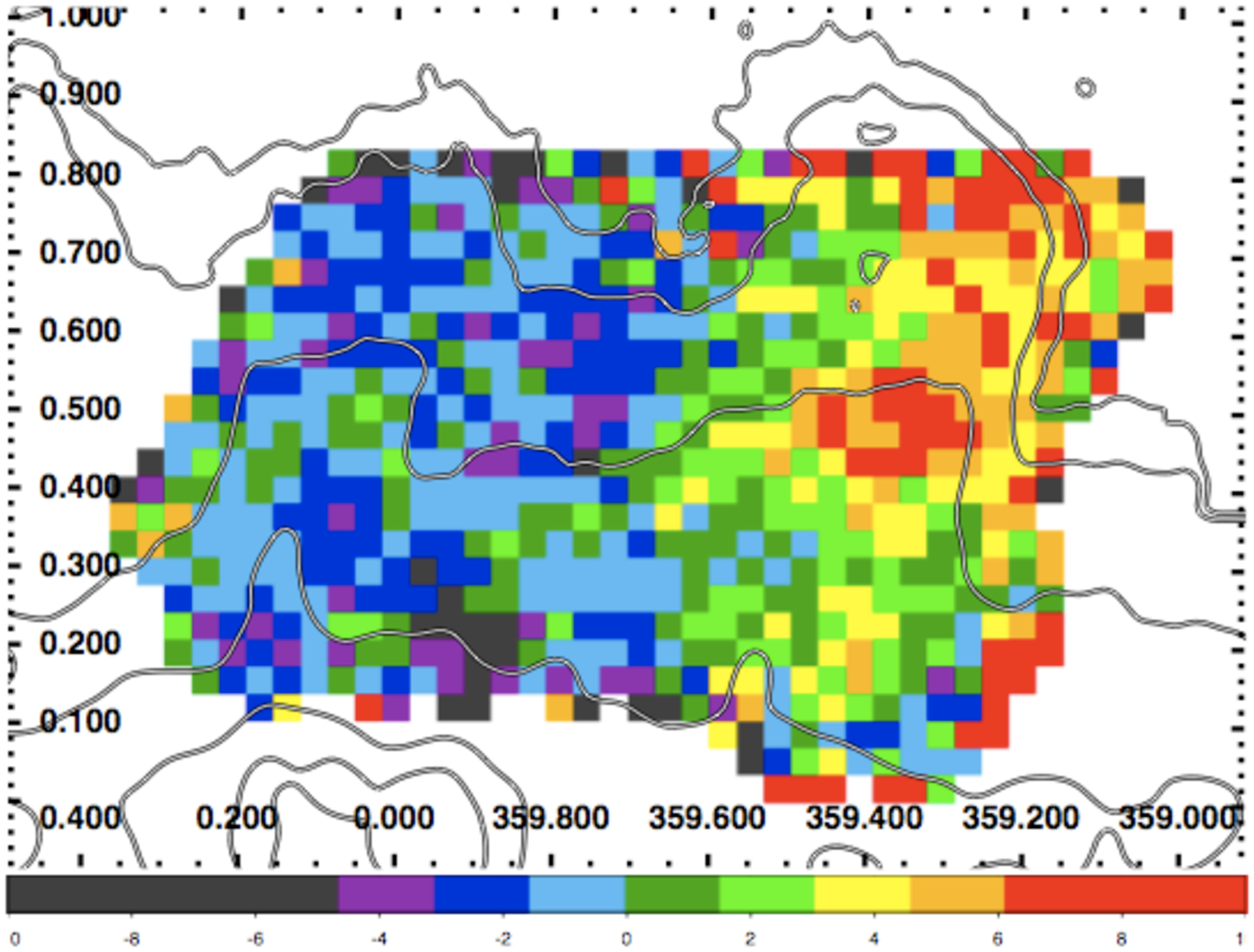}

\includegraphics[width=0.45\textwidth, trim=0 150 0 150, clip]{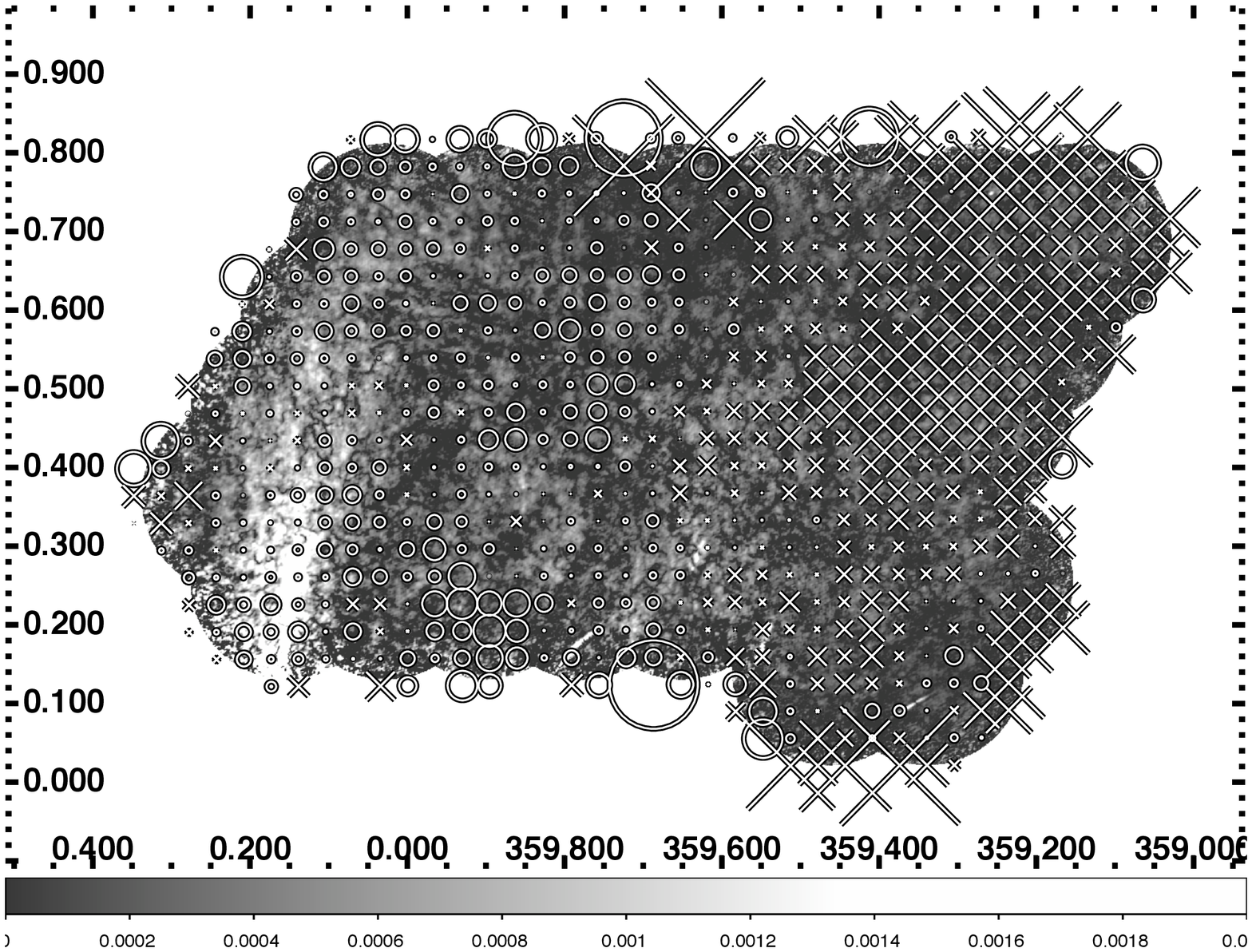}
\end{center}
\caption{\emph{Top}:  Map of \dt\ at 6 cm averaged over 125-arcsec tiles with the histogram-fitting method in units of degrees.  The contours show 6 cm brightness as observed by the GBT, similar to Fig. \ref{poln_polc}, with levels of $33*3^n$ mJy per 150\arcsec\ beam, with $n=0-5$. \emph{Bottom}:  Gray scale shows the 6 cm polarized intensity from 0 to 2 mJy, as in Fig. \ref{poln_polc}.  The symbols show the smoothed RM, assuming --220 rad m$^{-2}$ per \dt\ in the top panel.  Crosses show positions with RM$<0$ and circles showing RM$>0$. The size of the symbol is proportional to RM and ranges from --5720 to 4180 rad m$^{-2}$. \label{padilg}}
\end{figure}

Figure \ref{padilg} shows two maps of RM smoothed over 125-arcsec tiles with the histogram-fitting method.  The images show there is coherent structure on degree size scales.  The east side of the survey tends to have RM greater than zero and the west side less than zero.

\begin{figure}[tbp]
\begin{center}
\includegraphics[width=0.45\textwidth]{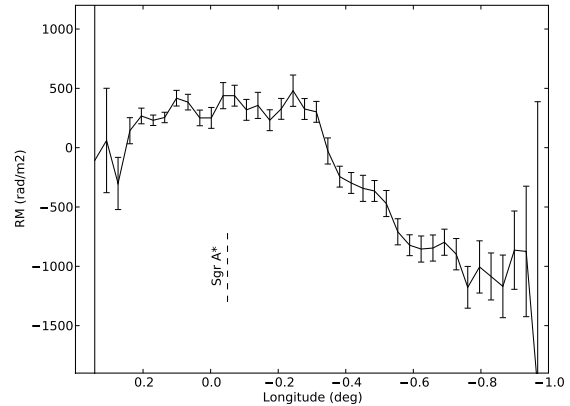}
\end{center}
\caption{The RM at 6 cm as a function of Galactic longitude, assuming 1\sdeg\ of \dt\ is equal to --220 rad m$^{-2}$.  The RM and its error are found by the histogram-fitting method for the entire latitude range of the survey ($0\ddeg1<b<0\ddeg8$) in strips of width 125\arcsec.  The longitude of Sgr A*, the dynamical center of the Galaxy, is shown with a dashed line.  \label{rmlong}}
\end{figure}

The east-west structure is seen more clearly in averages calculated over all latitudes, as shown in Figure \ref{rmlong}.  In the east, for $0\ddeg2<l<-0\ddeg3$, RM $\approx+330$ rad m$^{-2}$, then the RM changes rapidly for $-0\ddeg3<l<-0\ddeg55$, and in the west, for $-0\ddeg95<l<-0\ddeg55$, RM$\approx-880\pm50$\ rad m$^{-2}$.  Averaging RM over the top half of the survey ($0\ddeg45<b<0\ddeg7$) shows a similar structure as the average over all latitudes, but with a larger range.  The maximum RM is $+660\pm75$\ rad m$^{-2}$\ near $l=-0\ddeg25$\ on the east side, while the minimum RM is $-1320\pm75$\ rad m$^{-2}$\ near $l=-0\ddeg65$\ on the west side.

\subsection{Localized Features in the RM Image}
\begin{figure}[tbp]
\begin{center}
\includegraphics[width=0.4\textwidth, trim=0 130 0 150, clip]{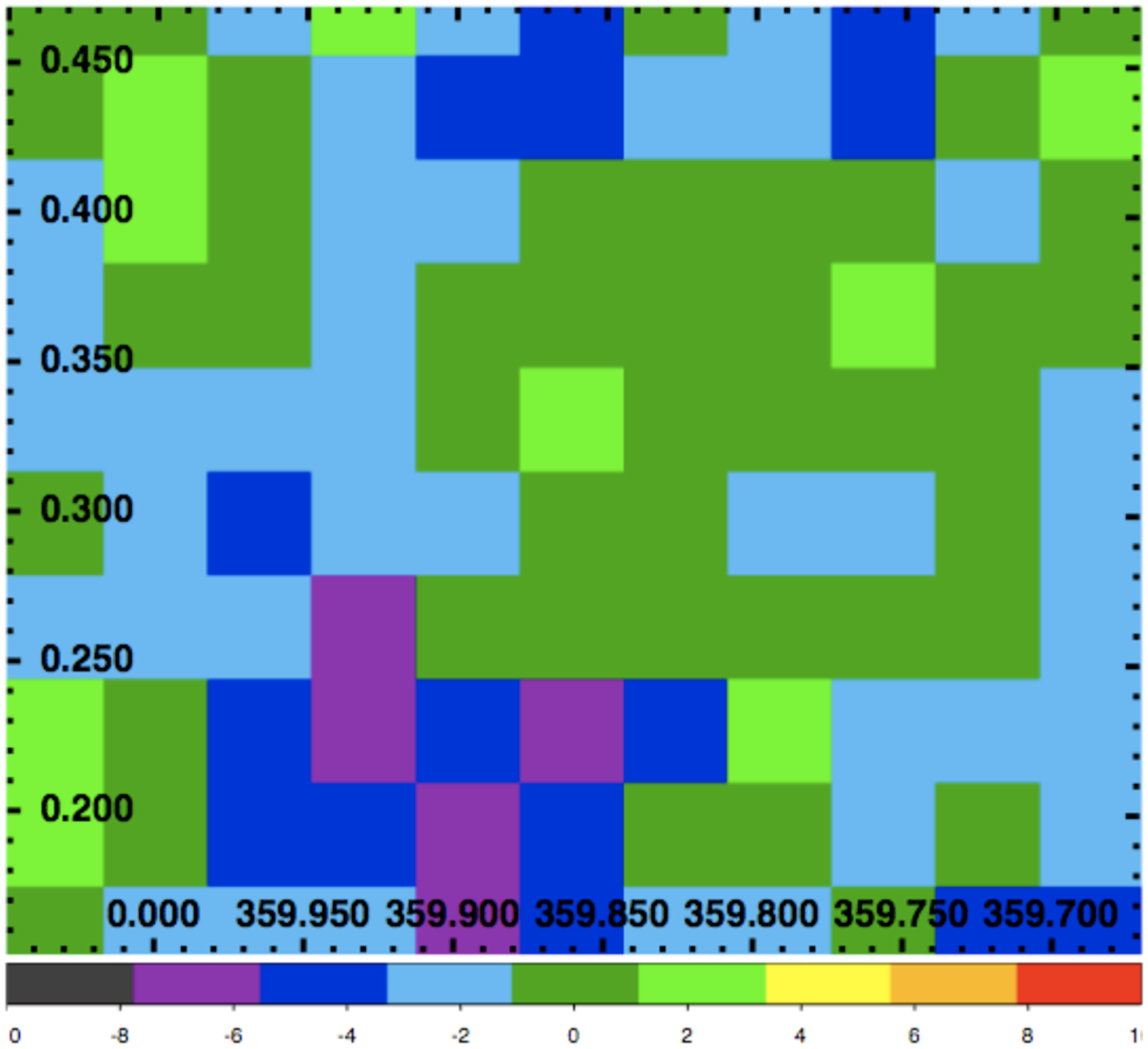}

\includegraphics[width=0.4\textwidth, trim=0 130 0 150, clip]{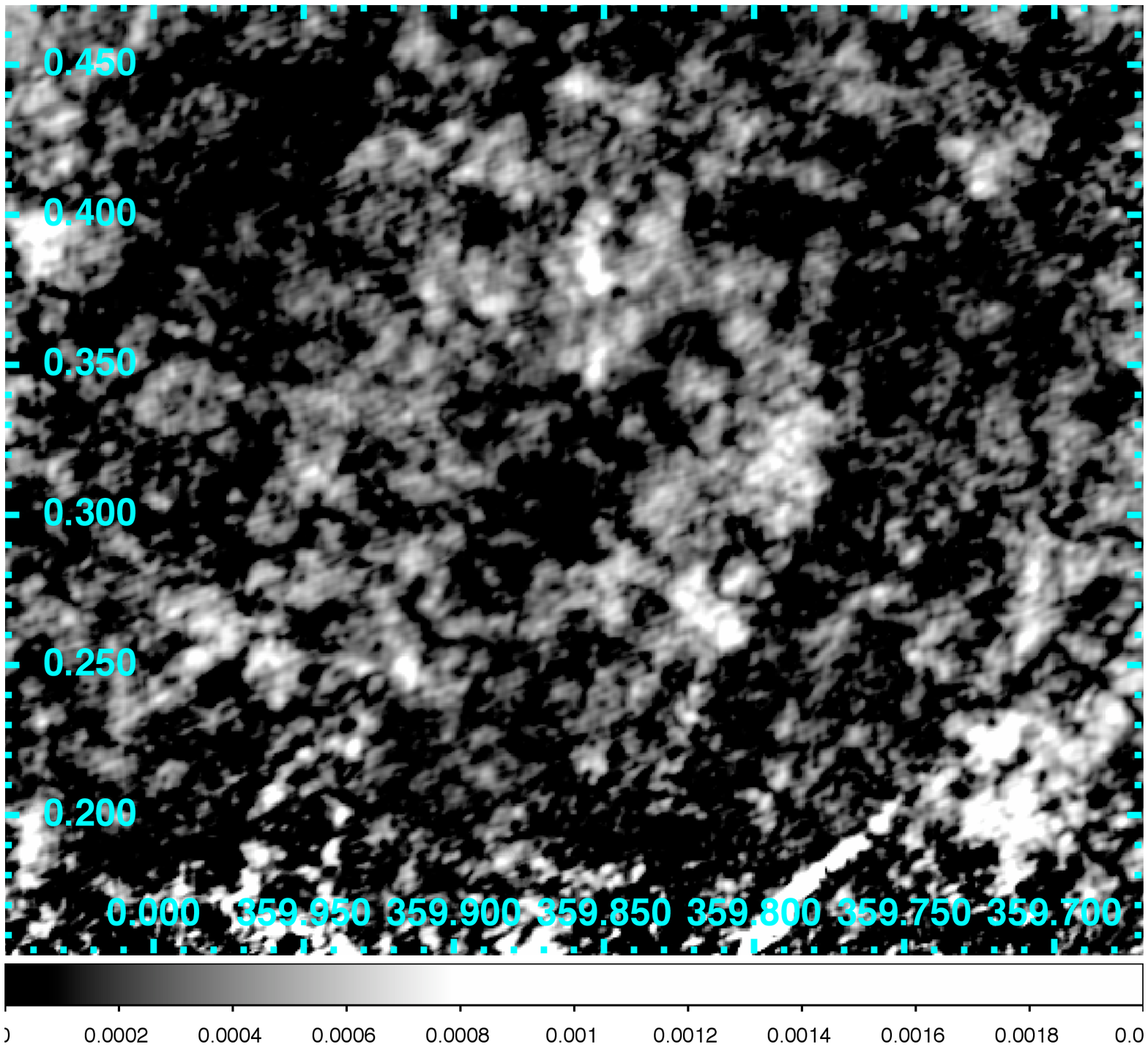}
\end{center}
\caption{\emph{Top}:  Smoothed map of \dt\ in units of degrees for the region near the region of large negative \dt.  \emph{Bottom}:  Map of the 6 cm polarized intensity for the same region as in the top panel.  The gray scale shows brightness ranging from 0 to 1 mJy beam$^{-1}$.  The NRF called the Cane \citep{l01} is located at the top of the loop at ($359\ddeg85$, $0\ddeg39$). \label{loop}}
\end{figure}

There are three arcminute-scale RM features that deviate from the simple structure described above.  One of the regions with the largest positive RM is at the southern border of the survey, near $(-0\ddeg1,0\ddeg2)$.  Figure \ref{loop} shows that the region with large RM covers a region about 8\arcmin\ across, just north of Sgr A.  The average RM for this region is $1188\pm198$ rad m$^{-2}$.  The average RM for all latitudes near $l=-0\ddeg1$ is $\approx+330\pm60$ rad m$^{-2}$.

The region with the most negative RM is at $(-0\ddeg6,+0\ddeg5)$, on the right side of Figure \ref{padilg}.  Over an area about 8\arcmin\ across, the mean RM is $-1320\pm110$ rad m$^{-2}$.  Figure \ref{rmlong} shows that the mean RM at $l\sim-0\ddeg6$ is $\approx-880\pm110$ rad m$^{-2}$.  This feature may partially explain why the amplitude of the east-west RM asymmetry is larger when averaging over the top half of the survey.

A third unusual RM structure is a ridge extending from $(0\ddeg25,+0\ddeg4)$\ to $(0\ddeg0,+0\ddeg5)$, seen at the left of Figure \ref{padilg}.  The feature has a negative RM, but the surrounding region has a positive RM.  The average RM along this ridge is $\approx-220\pm110$ rad m$^{-2}$, as compared to the mean value of $\approx330\pm60$ rad m$^{-2}$ for all latitudes near $l=0$\sdeg.  This structure is seen in the RM map of \citet{t86}, and a detailed comparison of that work to the present work is shown in Figure \ref{rmcomp}.  

\section{Discussion}
\label{poln_discussion}

\subsection{Modeling the Rotation Measure}
\label{modeling}
The complexity of the line of sight to --- and through --- the GC region argues for caution when interpreting RM patterns.  The apprent 6 cm RM suggests that the line of sight magnetic field changes sign, as if the field was predominantly azimuthal.  However, it is not immediately clear whether the observed polarized emission originates in the GC region or whether the observed RM can be used to measure properties of the GC magnetic field.  This section addresses these issues with modeling of the Galactic electron density and magnetic field.  We use the observed 6 cm RM longitude dependence to constrain parameters of the model and ultimately derive the expected polarimetric properties of the region over a range of wavelengths.

\subsubsection{Galactic Model}
The RM is calculated from a model of the electron density, $n_\mathrm{e}$, and the Galactic horizontal magnetic field, $\vec{B}$.  We use a cartesian Galactic coordinate system with the origin at $l=0$\sdeg\ at a distance of $r_\mathrm{GC}$\ from the Sun, the $x$\ axis pointing towards negative $l$, the $y$\ axis pointing away from the Sun, and the $z$\ axis pointing to positive $b$.  This technique does not calculate the emissivity of the synchrotron radiation, so we effectively assume that the polarized emission originates behind the Faraday rotating medium on the xz-plane.  The integral is done along a line to the GC distance, so the polarized emission is assumed to be at the peak of the electron distribution.  For this model, the Faraday depth along a given line of sight is:

\begin{equation}
\phi = \phi_0 + 0.81 \int_{r_\mathrm{GC}-5w}^{r_\mathrm{GC}} n_\mathrm{e}\vec{B}\cdot \mathrm{d}\vec{r},
\label{eqn:phi-model}
\end{equation}

\noindent where $\phi_0$\ is the foreground Faraday depth, $r$\ is the distance from the Sun along the line of sight, and $w$\ is the horizontal FWHM of the Galactic Center electron density enhancement \citep{c02}.  As described below, the Faraday rotation induced by the GC region dominates, so the limits of integration include only a path of length $5w$\ on the front side of the GC.  In \S \ref{depol}, we relax this assumption and consider the emissivity of the plasma.  The model electron density is:

\begin{equation}
\begin{array}{cc}
n_\mathrm{e}(x,y,z) = n_\mathrm{d} + & \\
n_{\mathrm{0}} \, \mathrm{exp} \left \{ -4\log{2} \left [ \frac{(x-x_0)^2 +(y-y_0)^2}{w^2}+ \frac{(z-z_0)^2}{h^2} \right ] \right \} , & \\
\end{array}
\end{equation}

\noindent where $n_\mathrm{d}$\ represents the electron density of the disc, excluding the Gaussian enhancement in the Galactic center. Here $n_\mathrm{d}$\ is assumed constant throughout the volume. The central enhancement is described by a three-dimensional Gaussian function, where $x_0$, $y_0$, and $z_0$\ are its offset, $h$\ is its vertical FWHM, and $n_0$\ is its maximum density.

Since the model is fit to RM measurements, the magnetic field model describes only the horizontal component.  The field points counter-clockwise as seen looking down on the plane, with the pitch angle $p$\ pointing slightly outward for positive $p$. The horizontal field strength is assumed to be a constant, $b_0$.

\begin{equation}
\mathrm{B}(x,y,z) = |b_0|\left(
\begin{array}{l}
- \sin (\theta-p)\\
\cos (\theta-p)\\
0
\end{array}
\right),
\end{equation}

\noindent where 

\begin{equation}
\theta = \tan^{-1}\frac{y-y_0}{x-x_0}.
\end{equation}

\subsubsection{Model Fit}
The 11 model parameters must be estimated carefully: there are only 39 longitude bins over a narrow $l$, $b$ range and only 2 frequency channels.  To simplify the procedure, we only solve for four parameters:  foreground RM, $x$-offset, magnetic field strength, and pitch angle.  The parameters $r_\mathrm{GC}$, $w$, $h$, $n_0$, $y_0$, and $z_0$ have default values taken from NE2001 \citep{c02}, while the disc electron density, $n_\mathrm{d}$, is the sum of the NE2001 thin disc and the \citet{g08} thick disc, evaluated at the Galactic center.  These default parameter values, shown in Table \ref{tab:parameter-values}, are consistent with other observations \citep{sp91,la05,l98,h06}.

\begin{figure}
\begin{center}
\includegraphics[width=\columnwidth, trim=0 0 50 0, clip]{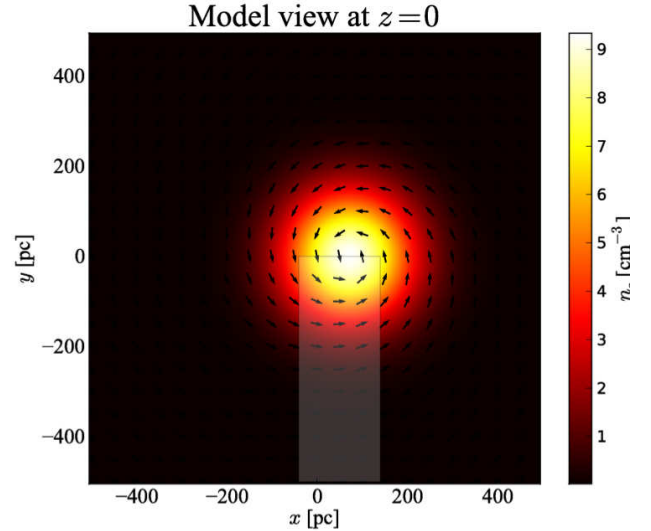}
\end{center}
\caption{Electron density and magnetic field in the plane of the Galaxy for the best-fit model.  For the least squares fit, the model was evaluated in the lightly shaded box at $b = +0\ddeg5$ or $z\approx+55$ pc. The model parameters are listed in Table \ref{tab:parameter-values}.}
\label{fig:model-view}
\end{figure}

Figure \ref{fig:model-view} shows a top-down view of the spatial distribution of the electron density and magnetic field.  The shaded region demonstrates that the observed area is rather small compared to the model structure.  This shows that the observations are limited to the center of the electron distribution.

We determine the best-fit parameter values and uncertainties with a three-step Monte-Carlo method. First, we randomly generate a list of 4\,000 mock observations.  The RM value at each $l$\ is drawn from a Gaussian distribution with mean and standard deviation equal to the RM and its error measured over the entire latitude range of the survey (as shown in Figure \ref{rmlong}).  Second, we fit Equation \ref{eqn:phi-model} for $b = +0\ddeg5$ to each mock data set by minimizing the reduced $\chi^2$ using the nonlinear, constrained, L-BFGS-B solver \citep{z94}.  Third, the mean and standard deviation of each parameter is measured from the ensemble of fit results.  

\begin{figure}
\begin{center}
\includegraphics[angle=90,width=\columnwidth]{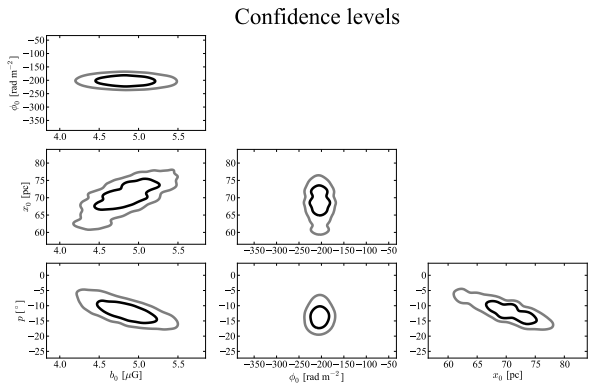}
\end{center} 
\caption{Confidence levels for all pairs of variable parameters. The 68\% confidence level contour is black, the 95\% confidence level contour gray.}
\label{fig:confidence}
\end{figure}

The parameters and fit results for the longitude dependence of the 6 cm RM are summarized in Table \ref{tab:parameter-values}.  Figure \ref{fig:confidence} shows the 68\% and 95\% confidence levels for all pairs of variables. The foreground RM is orthogonal to all other parameters, while the remaining three show mild degeneracies.  
 
\begin{table*}[htb]
\caption{\label{tab:parameter-values} Model parameters}
\begin{center}
\begin{tabular}{lrrlll}
\hline
\hline
 Name             &  Value  &  $\pm$  &  Unit          &  Range\tablenotemark{a} &  Description                       \\
\hline
 $b_0$            &   4.84  &   0.25  &  $\mu${}G      &  $[0.1, 20]$       &  Magnetic field strength in plane   \\
 $p$              &  -11.6  &    3.9  &  $^\circ$      &  $[-25,25]$         &  Pitch angle                       \\
 $\phi_0$         &   -210  &     44  &  rad m$^{-2}$   &  $[-1000, +1000]$  &  Foreground RM                     \\
 $x_0$            &     70  &      3  &  pc            &  $[0,100]$         &  Offset in $x$                     \\
 $y_0$            &      0  &         &  pc            &                    &  Offset in $y$                     \\
 $z_0$            &    -20  &         &  pc            &                    &  Offset in $z$                     \\
 $n_0$            &     10  &         &  cm$^{-3}$     &                    &  Peak of GC electron density Gaussian  \\
 $n_\mathrm{d}$    &  0.019  &         &  cm$^{-3}$     &                    &  Electron density of disc          \\
 $r_\mathrm{GC}$   &   8000  &         &  pc            &                    &  Distance to GC                    \\
 $w$              &    240  &         &  pc            &                    &  Horizontal FHWM                   \\
 $h$              &    125  &         &  pc            &                    &  Vertical  FHWM                    \\
\hline
\tablenotetext{a}{The range of parameter values allowed during the constrained nonlinear model fit.  If no range is shown, the parameter is fixed.}
\end{tabular}
\end{center}
\end{table*}

\begin{figure}
\begin{center}
\includegraphics[width=\columnwidth]{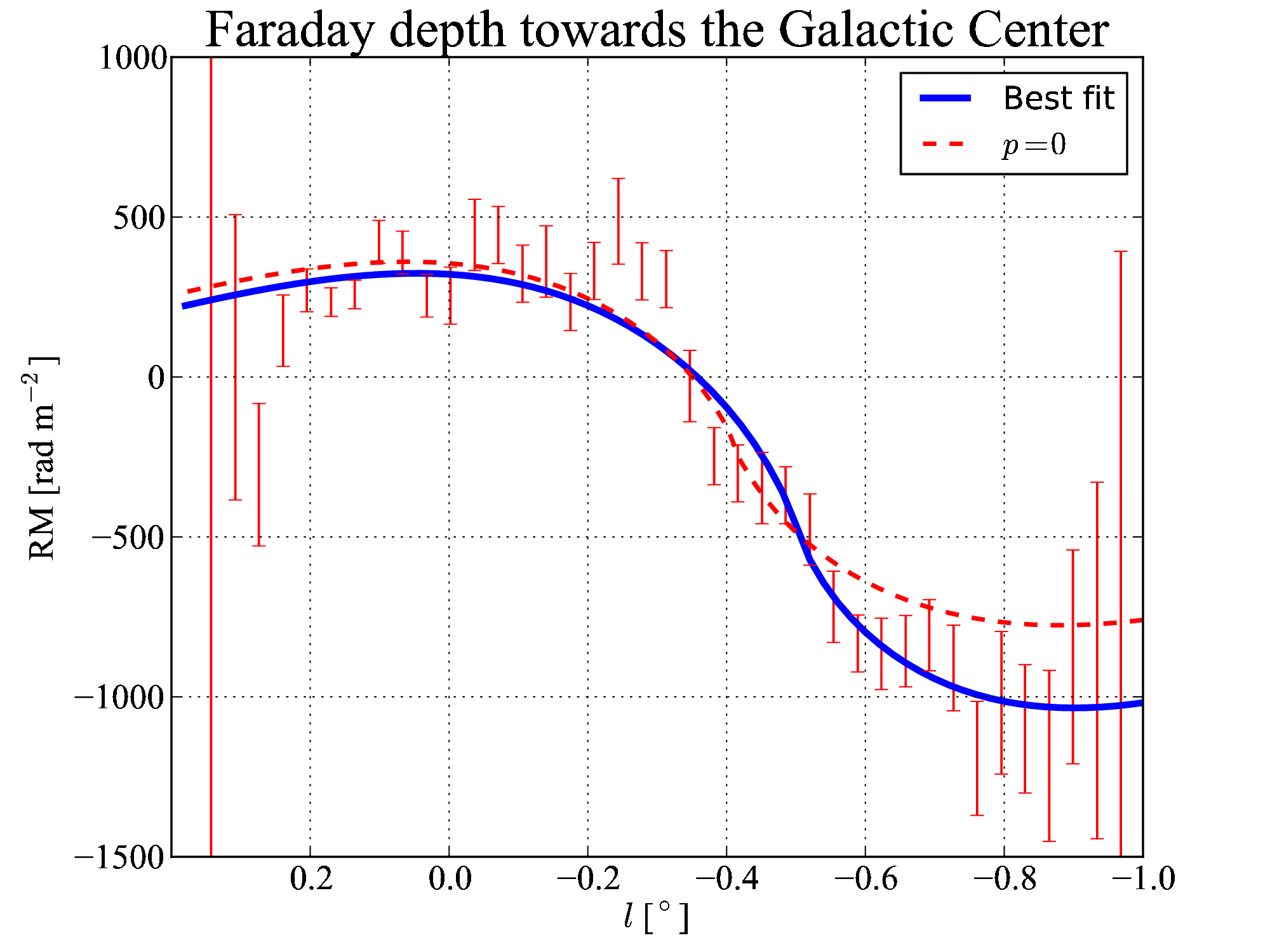}
\end{center}
\caption{The 6 cm RM as a function of Galactic longitude in the GC region.  The solid blue line shows the observed RM, measured over the entire latitude range of the survey ($0\ddeg1<b<0\ddeg8$).  The dashed red line shows the best-fit model at the mean latitude of the survey ($+0.5^\circ$; parameters shown in Table \ref{tab:parameter-values}).
\label{fig:faraday-depth}}
\end{figure}

The 6 cm RM predicted by the best-fit model is compared to the observed RM in Figure \ref{fig:faraday-depth}.  Since the geometry of the magnetic field is not well known in the GC region, we compare the observations to the best-fit model with and without a pitch angle parameter.  For the model with the pitch angle (and shown in Table \ref{tab:parameter-values}), the reduced $\chi^2$ is 1.6 with 34 degrees of freedom.  The model with no pitch angle parameter is slightly worse, particularly at $l < -0.6^\circ$, with a reduced $\chi^2$ of 2.2 and 35 degrees of freedom.  A $p$ of 0 is ruled out formally at the 3 $\sigma$ level, although as discussed in \S \ref{depol} considering emissivity of the model will change some details of the best-fit parameters.

The quality of the fit shows that a realistic GC electron distribution and magnetic field geometry can explain the observed RM longitude pattern.  In particular, the rapid change of RM with Galactic longitude is best explained by the rapidly changing magnetic field orientation within the central few hundred parsecs.  This is consistent with our assumption that most of the polarized emission originates in the center and that the Faraday rotation happens on the near side of the GC.

Not only can the model fit the observed RM pattern, but the best-fit values are consistent with other observations.  The magnetic field strength and electron density are degenerate in the model, but we constrain the horizontal component of the magnetic field to be roughly $5 (10\ \rm{cm}^{-3}/n_e^{\rm{gc}}) \mu$G at heights of about $0\ddeg5$.  Considering the predominantly vertical orientation of the magnetic field \citep{n10}, the implied total field strength is consistent with that measured previously \citep{h92,f09,c10}.  The foreground RM is similar in magnitude to that observed toward the pulsars in the inner Galaxy, which have RM up to $\sim$200 rad m$^{-2}$ \citep{ma05,h06}.  We compare this model to other measurements of the foreground RM in \S \ref{gcmag}.

The physical consistency of the model to GC region argues for a GC origin of the polarized emission.  The shape of the RM distribution requires a value of $n_e \cdot B_{||}$\ that exists only in the GC region \citep{l98}.  Also, the typical $n_e$ and $B_{||}$ in the Galactic plane are not changing enough with longitude to cause the observed RM pattern.

\subsubsection{Faraday Depth Estimate}
\label{depol}
While the model is based on the observed 6 cm RM, reasonable assumptions about the synchrotron emission in our volume will allow us to predict RM and depolarization at all frequencies and positions.  These assumptions are not parameterized in our initial model fit, so its predictions can comment on the reasonableness of our model.  Below we show how the model predicts a Faraday depth distribution and how it affects the interpretation of the observed polarization at 6 cm.

The following treatment of synchrotron radiation closely follows that of \citet{s08}.  The synchrotron intensity of a slice with thickness $d\vec{r}$\ is given by

\begin{equation} 
I\ d\vec{r} \propto n_\mathrm{rel} \vec{B_\perp}^{2}\nu^{-1}\ d\vec{r},
\end{equation}

\noindent where $n_\mathrm{rel}$\ is the relativistic electron density.  We assume that $n_\mathrm{rel}$\ is constant in the inner 3 kpc of the Milky Way, that the intrinsic fractional polarization of each volume element is constant in frequency and space, and that the shape of the synchrotron spectrum is the same in all volume elements. We therefore drop the frequency dependence and relativistic electron density from our simulation of the polarized intensity:

\begin{equation}
fI\ d\vec{r} \propto \vec{B_\perp}^{2}\ d\vec{r},
\end{equation}

\noindent where $f$\ is the fractional polarization. If the magnetic energy density is equal to the gas energy density, then, according to \citet{m04},

\begin{equation}
\|\vec{B}\| \propto n_\mathrm{e}^{1/2}.
\label{brentjens_abell2256_eqn:b_propto_ne}
\end{equation}

\noindent Because $\|\vec{B}\|$\ is likely an order of magnitude stronger than the horizontal field \citep{c10}, we assume for the sake of simplicity that $B_\perp\approx \|\vec{B}\|$, and therefore

\begin{equation}
fI\ d\vec{r} \propto n_\mathrm{e}\
d\vec{r}.
\label{eqn:diff-fractional-polarization}
\end{equation}

We can now combine the polarized emissivity from Equation \ref{eqn:diff-fractional-polarization} with the Faraday depth and Faraday thickness to create the Faraday dispersion function $F(\phi)$.  The dispersion function is the polarized flux as a function of Faraday depth, which can be Fourier transformed into the complex fractional polarization as a function of $\lambda^2$ \citep{br05}.  We assume that the intrinsic polarization angle is the same thoughtout the GC region, which is reasonable given the large-scale organization reported elsewhere \citep{n10}.  The integration was done from $r_\mathrm{GC}-5w$\ to $r_\mathrm{GC}+5w$ to predict the emission across the entire GC region.  All lines of sight are normalized to have a polarization fraction of 70\% at $\lambda^2 = 0$.

\begin{figure*}
\begin{center}
\includegraphics[width=\textwidth]{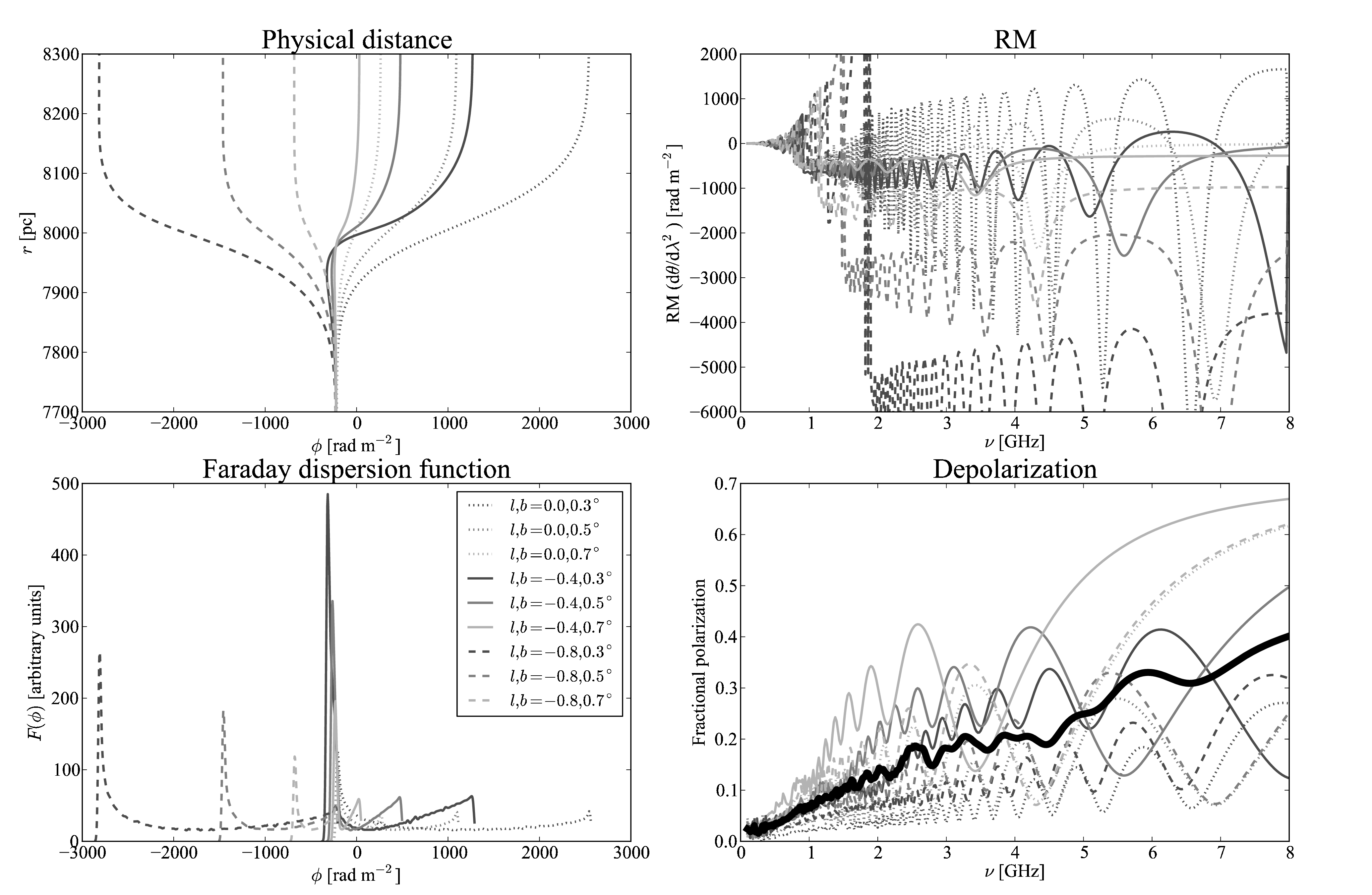}
\end{center}
\caption{Observational predictions of the model. The legend at the bottom left is valid for all panels. \emph{Top left}:  Physical distance as a function of Faraday depth. \emph{Bottom left}: Faraday dispersion function at same $\phi$ scale as top left. \emph{Top right}: Slope of polarization angle versus $\lambda^2$. \emph{Bottom right}: Fractional polarization. The thick curve is the average of the other curves.}
\label{fig:depolarization}
\end{figure*}

Figure \ref{fig:depolarization} shows several predictions of the model given the assumptions described above.  Since the initial model was fit assuming emission at the GC distance and RM induced only in the foreground of the GC (i.e., \ref{eqn:phi-model}), we expect this model to be more Faraday thick.  Indeed, while the underlying model is smooth and simple, the RM and polarization fraction are highly structured as a function of latitude and frequency.  It is also worth noting that the predictions of the model are idealized in the sense that they do not account for beam depolarization or finite bandwidth of the observations.  As such, they likely overpredict the polarization fraction and frequency-dependent changes in RM.

The relation between Faraday depth and physical distance is displayed in the top left panel of Figure \ref{fig:depolarization}.  The Faraday depth changes little far from the GC because the medium is tenuous.  However, it also changes little in dense regions where the field direction reverses, such as in front of and behind the GC.  As a result, a significant amount of flux will end up at the Faraday depths of the vertical sections of the $r$ versus $\phi$ plot.  The flux \emph{per unit $\phi$} is large in these regions, as shown in the Faraday dispersion plot in the bottom left panel.  Because these caustic-like features occur whenever the line-of-sight field reverses in a synchrotron emitting area (Ue-Li Pen, 2010, private communication), RM synthesis of Faraday thick areas is much more sensitive to these reversals than to the bulk emission at large $\phi$ scales.

If we think of the polarized emission as a complex Stokes vector that rotates according to its Faraday depth, we can imagine the effect on the observed RM.  The peaks in $F(\phi)$ interfere to create complex Faraday effects as a function of $\lambda^2$.  As shown in the top right panel, the RM as determined by observing only two nearby frequencies can vary widely and argues for caution when interpreting our physical model.  This shows that considering the emission from the entire GC region (not just the Faraday rotation by the foreground) makes the region Faraday thick.  

\begin{figure*}
\begin{center}
\includegraphics[angle=90,width=\textwidth, trim=50 0 0 0, clip]{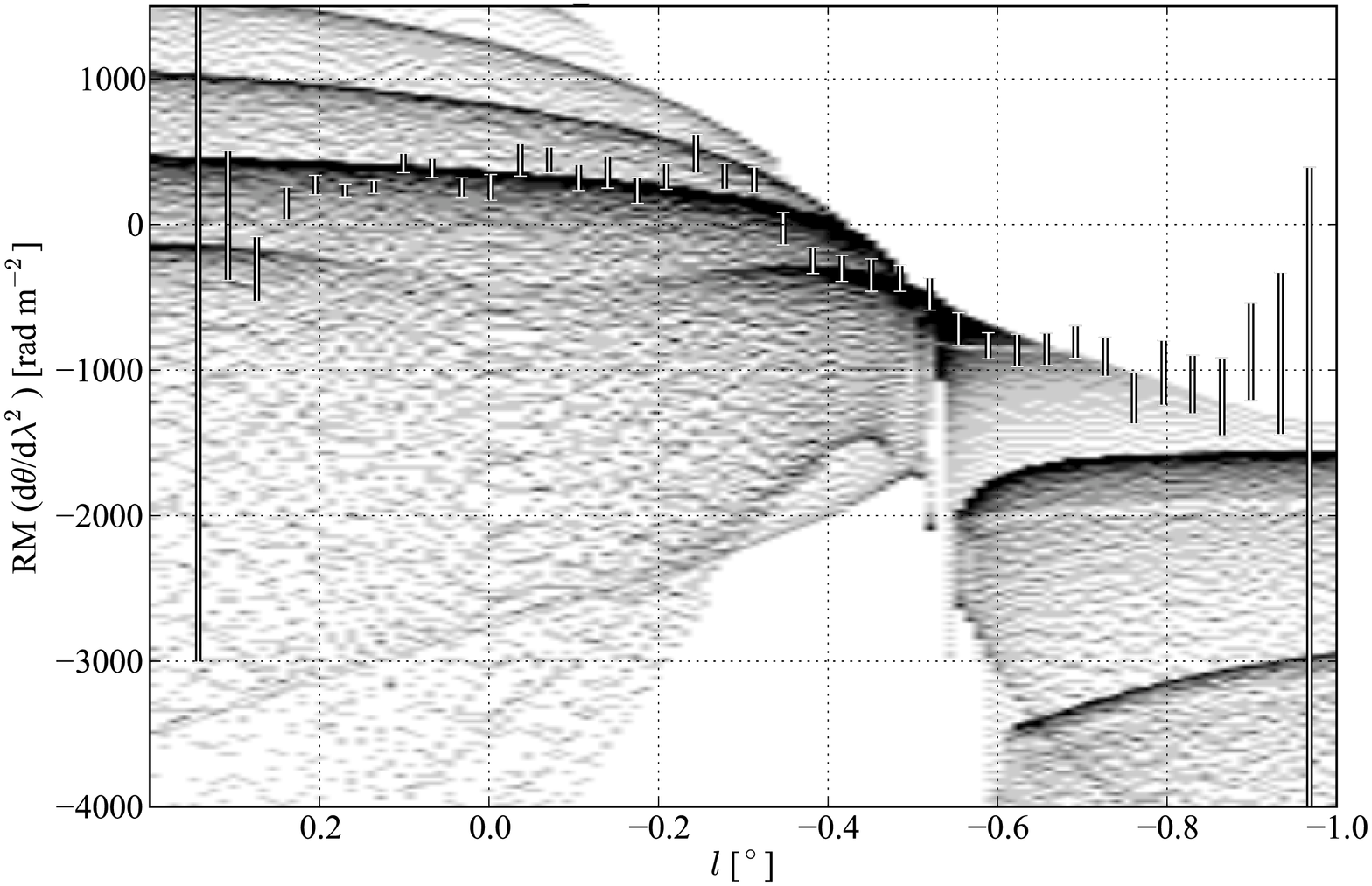}
\end{center}
\caption{Comparison of the predicted and observed RM at 4.86 GHz (near 6 cm) as a function of Galactic longitude.  The gray scale shows RM values expected for lines of sight with b ranging from 0\ddeg2 to 0\ddeg7.  The prediction was made using the model shown in Tab. \ref{tab:parameter-values} and considering the emission and Faraday rotation though the entire GC region.  The vertical bars show the observed RM at 4.86 GHz, as shown in Fig. \ref{rmlong}.
}
\label{rmhist}
\end{figure*}

Despite the Faraday thickness, the model shows that the RM at 6 cm preserves the observed east-west gradient.  Figure \ref{rmhist} compares our observed RM at 6 cm to the RM derived along many lines of sight through this model.  This is similar to the RM derived from the model in Figure \ref{fig:faraday-depth}, which did not calculate the Faraday dispersion function, but instead assumed Faraday rotation occurred in the foreground.  Figure \ref{rmhist} confirms that considering the emissivity produces a wide range of RM, but that there is a clear east-west gradient. \footnote{The observed RM matches the model in the east, but tends to be more positive than the model in the west.  This shows that considering the emissivity and the full Faraday thickness requires more detailed modeling in the western part of the survey region.} The qualitative agreement between models with and without emissivity shows that the observed RM gradient is caused by the orientation of the magnetic field in the GC region.

Figure \ref{fig:depolarization} also shows the polarization fraction expected when considering the emissivity of the model.  As we assume an intrinsic polarization fraction of 70\%, the plot shows that nearly every line of sight will have significant depolarization.  On average, the predicted polarization fraction is 20\% at 4.8 GHz and 7\% at 1.4 GHz.  Since the simulation does not include beam depolarization and assumes a perfectly organized magnetic field, the predictions should overestimate the polarization fraction.  We consider the predictions in agreement with the observed polarization fraction of 10 to 20\% at 4.86 GHz and $<1$\% at 1.4 GHz.

In summary, considering the emissivity of our best-fit model for the GC magnetized plasma shows that the observed RM structure has a complex connection to the actual physical properties.  However, the RM trend with Galactic longitude is directly related to the magnetic field direction in the GC region.  More detailed physical modeling will require polarimetry with hundreds of channels between 2 and 8 GHz, such as with the EVLA \citep{u06}.

\subsection{Coincidence of RM for Extended and Filamentary Emission}
\label{rmdisc}
Eight NRFs were detected in polarized emission at 6 cm and seven of these have reliable RM measurements \citep{gcl_vla}.  Interestingly, all but one of these have RM consistent with their surrounding diffuse emission.  In other words, the RMs toward the NRFs largely follow the longitude dependence found toward the diffuse emission.  If the RMs toward the filaments was unrelated to that of the diffuse emission, a binomial probability distribution predicts a 5\% chance of 6/7 coincidences.

The similarity of the RMs in the diffuse and filamentary emission is consistent with observations of the brightest NRF, the Radio Arc.  As shown in \S\ \ref{poln_comparison} and elsewhere \citep{y88}, the morphology and RM measured toward the Radio Arc has a continuous connection into the polarized diffuse emission in the east of this survey.  Since the Radio Arc is known to be within the central 100 pc of the GC \citep{y87,l99,l89}, some of the diffuse polarized emission must also be located in the GC.

The fact that the filaments and the diffuse polarized emission are physically near each other could explain their similar RMs.  The simplest model to explain this coincidence is that the filaments and diffuse emission are behind the same Faraday screen.  According to our modeling, such a screen would be located within the central kiloparsec.  However, NRFs are known to have RM changes that coincide with physical changes, which argues that some RM is induced locally \citep{la99}.  A locally induced RM is consistent with the strong magnetic fields inferred in NRFs \citep{y97}.

If this RM coincidence is not extrinsic, then it must be intrinsic:  the diffuse emission and the NRFs have similar magnetic field orientations.  Physically, this can be explained if the NRFs are local enhancements and perturbations of the global magnetic field.  This concept is common to several models for generating NRFs \citep[e.g.,][]{b88,s94,b06}.  However, it excludes the model of \citet{s99}, which relies on an interaction of molecular clouds with a global wind to generate the NRFs.  Other models \citep[e.g.,][]{r96} do not clearly predict how filaments can enhance and perturb the global magnetic field.

The NRFs and the diffuse polarized emission also have a similar spatial distribution.  As described earlier, the RM observed in the diffuse emission changes sign near $l\approx-0\ddeg35$.  Other observations, described in \S \ref{gcmag}, show that most RM measurements in the GC region follow a pattern that is centered near $l\approx-0\ddeg35$.  High-resolution 20 cm survey of \citet{y04} showed that there are dozens of candidate NRFs in the GC region with the highest density between $l=0\ddeg2$ to $-0\ddeg7$.  This shows that the center of the distribution of NRFs is similar to the center of the RM pattern.  A third coincidence is that the GC lobe, a shell of gas related to a mass outflow, is centered near the same longitude.  These coincidences indicate a connection between the GC lobe outflow and the GC magnetic field.

\subsection{Degree-Scale Structure in the GC Magnetic Field}
\label{gcmag}
\begin{figure}[tbp]
\begin{center}
\includegraphics[width=0.5\textwidth]{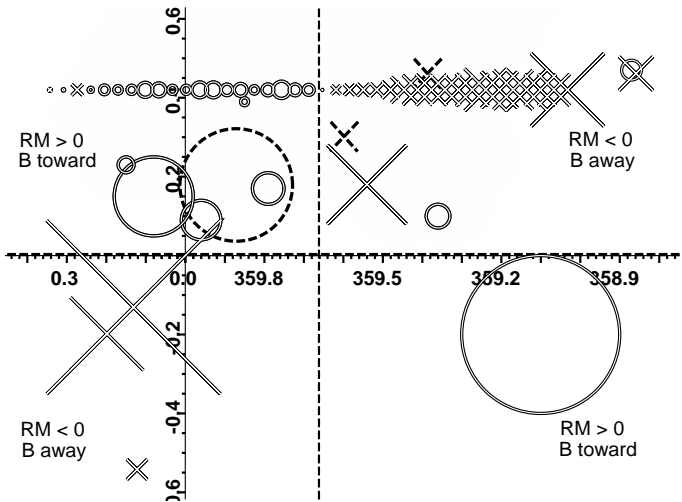}
\end{center}
\caption{Schematic diagram of measurements of RM in the central degree of the Galaxy.  The cross symbols show sources with negative RM and circles show sources with positive RM. The size of the symbol is proportional to the value of RM, with values ranging from 1100 to --2200 rad m$^{-2}$.  The horizontal series of symbols show the RM averaged over all latitudes of the present survey.  The other solid-lined symbols show measurements from the literature, specifically:  G0.08+0.15 \citep[a.k.a. ``the Northern thread''][]{l99}, the Radio Arc \citep[south of $b=0$\sdeg;][]{t86}, G0.87--0.87 \citep{re03}, G359.1--0.2 \citep[a.k.a. ``the Snake'';][]{g95}, G358.85+0.47 \citep[a.k.a. ``the Pelican'';][]{la99}, and new filaments described in \citet{gcl_vla}.  Note that the two symbols furthest to the top-right represent the max and min RM measured toward the unusual ``Pelican'' NRF.  The dashed crosses and circles show the RM measured toward extragalactic sources \citep{r05}.  The dashed horizontal and vertical lines split the region into quadrants that mostly have similar RM values.  \label{poln_rmallschem}}
\end{figure}

Figure \ref{poln_rmallschem} shows a schematic of all RM measurements toward sources believed to be in or beyond the GC region.  The RM values derived earlier in this work are shown along with observations of the extended polarized emission from the Radio Arc \citep{t86}, other NRFs \citep{la99,l99,g95,re03}, and background sources \citep{r05}.  North of the plane, the east-west gradient in RM is seen toward diffuse, compact, and filamentary sources.  The pattern is antisymmetric about $l\approx-0\ddeg35$ and across the Galactic plane \citep[esp. ][]{t86}.  

The simplest pattern to describe the GC RM values is that of a checkerboard (four quadrants of alternating sign) shifted $0\ddeg35$ ($\approx$50 pc) west of the center.  The possibility of a checkerboard pattern centered at $l=0$\sdeg\ has been noted before \citep{u85,n03}, but some RM values were not consistent with the pattern \citep{f09}.  The RM reported in this work finds that the checkerboard pattern is robust if it is assumed to be shifted from the center.

The checkerboard pattern inspired the ``flux-dragging'' model for the GC magnetic field \citep{u85, n03}.  The model explains a large-scale pattern in the GC RM as the effect of Galactic rotation on a frozen-in, poloidal (vertical) magnetic field.  As the disk rotates, the magnetic field in the disk is dragged away from us on the east side and toward us on the west side.  This pull creates a line-of-sight component of the magnetic field.  This perturbation to the magnetic field has a checkerboard pattern in the sign of the RM, such that the RM will have opposite signs toward any two adjacent quadrants formed about the center of rotation.  

Interestingly, the parity of the checkerboard pattern constrains the orientation of the magnetic field, breaking the 180\sdeg\ ambiguity of observations of the polarization angle.  If this scenario is valid, the parity of the observed pattern is consistent with the magnetic field pointing from south to north.  This is the only known measurement that can break this ambiguity in the orientation of the magnetic field on the plane of the sky.

\citet{r08} found that the flux-dragging scenario was inconsistent with the RM observed toward extragalactic sources seen through the central 12\sdeg.  The RM toward their 60 background sources had an average RM$=+413$\ rad m$^{-2}$ and no checkerboard pattern.  While the the RM values over the central 12\sdeg\ are predominantly positive, the three measurements in the central 2\sdeg\ studied here (G359.388+0.460, G359.604+0.306, G359.871+0.179) agree with the shifted checkerboard pattern.  

However, there are clear differences between the RM structure seen here and that reported by \citet{r08}.  For example, the average RM observed by \citet{r08} in the central 12\sdeg\ is not consistent with the foreground RM in our model ($\phi_0$).  Assuming that half of the average RM is contribued by the foreground to the GC, we'd expect $\phi_0 = 206.5$\ rad m$^{-2}$, but find roughly the opposite of that.  One possibility is that the foreground RM fit by our model is poorly constrained by our data.  Indeed, \S \ref{depol} shows that the absolute level of RM is poorly constrained by these narrow-band observations, but that relative values are well constrained.  Another possibility is that the central 2 degrees of the Galaxy is magnetically and dynamically different from the region beyond, making comparison with the RM measured in \citet{r08} less meaningful.  The presence of NRFs \citep{n04,y04} and the ``central molecular zone''\citep{m96} makes the central 2\sdeg\ notably different from the region beyond.

In summary, we argue that RM measured in the central 2\sdeg\ of the Galaxy is distinct from RM structure seen outside this region and that it reflects large-scale GC magnetic field structure.  The RM structure is simplest to describe as a checkerboard pattern shifted $\sim50$\ pc from the dynamical center of the Galaxy.  The longitude shift requires some non-Keplerian motion of the ionized gas in the GC region.  This is consistent with recent observations showing that the GC is host to a small starburst outflow centered $\sim50$\ pc west of the GC \citep[the GC lobe;][]{b03,gcl_all}.  The distribution of NRFs, a more direct tracer of the GC magnetic field, is also centered tens of parsecs west of the GC \citep[see Figure 29 of ][]{y04}.  It is clear that the electron and magnetic field distributions are not symmetric about the $l=0$\sdeg;  our new RM measurements confirm this.


\section{Conclusions}
We have presented observations and modeling of polarized 6 cm radio continuum emission toward 0.5 square degrees of the GC region.  The radio continuum survey detects polarized emission thoughout the region in the form of diffuse polarized emission, compact sources, and filamentary sources.  The two bands in the continuum observations allow us to measure RM to this polarized emission.  We develop a statistical technique to measure the RM;  comparing our results to more robust RM measurements shows that our technique is reliable.

There is a striking large-scale pattern in RM toward the diffuse polarized emission.  Values in the eastern part of the survey are generally about +330 rad m$^{-2}$, but change to $-880$\ rad m$^{-2}$\ in the western part of the survey.  There is a sharp transition around $l=-0\ddeg35$\ at all latitudes in the survey.  Modeling of the propagation of the polarized signal shows that this pattern is induced within $\sim$1 kpc of the GC region.  The RM measured toward radio filaments known to be in the GC region are generally consistent with that of the diffuse polarized emission.  This coincidence is consistent with models for the filaments as localized enhancements to a global magnetic field.

The modeling of the GC magnetized plasma shows that the RM structure constrains the orientation of the GC magnetic field.  This RM pattern shows that the GC magnetic field is organized on size scales of roughly 150 parsecs.  Combining these and other RM measurements in the GC region, we strengthen earlier suggestions for a checkerboard pattern in RM covering the central 300 parsecs, but only if the structure is shifted roughly 50 pc west of the dynamical center of the Galaxy.  We show that the RM measured along different lines of sight and toward different tracers are consistent with this shift.

The observed polarization and RM in the GC is consistent with the GC having a poloidal magnetic field that is perturbed by the motion of gas in the Galactic disk \citep{u85, n03}.  This model is being supported by a growing body of evidence \citep{c03,n10}.  Under this model, our RM observations constrain the GC magnetic field to be directed from south to north.  Our observations also suggest that a second-order perturbation, a small outflow from the GC, has shifted the magnetic symmetry axis of the GC about 50 pc west of the dynamical center of the Galaxy.

New observations can test this model in several ways.  First, observing the diffuse polarized emission between 6 and 20 cm (5 and 1.4 GHz), where it becomes Faraday thick, would constrain models of its physical distribution.  Expanded VLA observations with thousands of channels at these frequencies will track the Faraday rotation and depolarization well enough to create a 3D reconstruction of the magnetic field topology in the Galactic Center.  Second, measuring the RM of other radio filaments would test the idea that they are preferentially aligned with the RM of the extended polarized emission.  Third, the detection of diffuse polarized emission and its RM beyond the region studied here (particularly south of the plane) would confirm that it traces a general property of the GC region.

\acknowledgements{We thank Farhad Yusef-Zadeh, Bryan Gaensler, Bill Cotton, and Dominic Schnitzeler for valuable discussions during this work.}

{\it Facilities:} \facility{VLA}

\begin{appendix}

\section{Spatially Smoothing \dt}
\label{meandt}
\subsection{Mean \dt\ Images}
To improve visualization of the \dt\ images, the mosaics were smoothed using two independent methods.  Both methods were implemented in IDL.  The first method of smoothing \dt\ was to measure the error weighted mean value of \dt\ over small regions.  Hereafter, we refer to this as the mean method.

A major caveat to this method is that averaging angles is not proper, since they represent vector quantities.  However, averaging angles is approximately correct for small angles, which is usually true for the \dt\ images presented here (Fig. \ref{histcomp}).  Furthermore, the large values in the \dt\ map are generally in the noisiest parts of the image and are down-weighted by large errors.  There was no significant difference when using the median instead of the mean.

\subsection{Fitting Histograms of \dt\ Images}
A second, more robust method of smoothing the \dt\ images is to fit the distribution of \dt\ values with a model.  Since a model is fit to the histogram, this approach avoids the problems of averaging angles.  The model can also parameterize the noise- and signal-like contributions to the \dt\ distribution, making it robust to outliers.  

As described in \S \ref{padianalysis}, histograms of \dt\ were fit with a Lorentzian plus a constant background.  The best-fit values and their errors were found by the Levenberg-Marquardt algorithm, implemented as MPFIT in IDL \citep{m09}.  All errors reported are 1$\sigma$ confidence intervals.  

Figure \ref{histcomp} shows two examples of histograms extracted with best-fit models. A histogram bin size of 2\ddeg5 was used;  varying the bin size by a factor of a few does not significantly change the fit. The smoothed maps have pixel sizes of 125\arcsec$\times$125\arcsec, which are small enough to resolve structure in the mosaic, but large enough for accurate fit results.  

An advantage of the histogram-fitting method is that it calculates the mean and error in \dt\ from the distribution of \dt; no noise image is required.  A disadvantage of the histogram-fitting method is that it assumes that all pixels in the region sampled are a part of the same distribution.  In fact, there are times when multiple sources, with different distributions of \dt\ (i.e., different RMs) are sampled by the same histogram.  In this case, the source that occupies the most pixels will dominate the histogram distribution and the best-fit value of \dt.  This is different from what happens when calculating an error-weighted average of \dt, in which the average is dominated by the pixels with the lowest noise (i.e., the brightest sources).

\subsection{Comparing Mean and Histogram-Fitting Methods}
\begin{figure}[tbp]
\begin{center}
\includegraphics[width=0.8\textwidth]{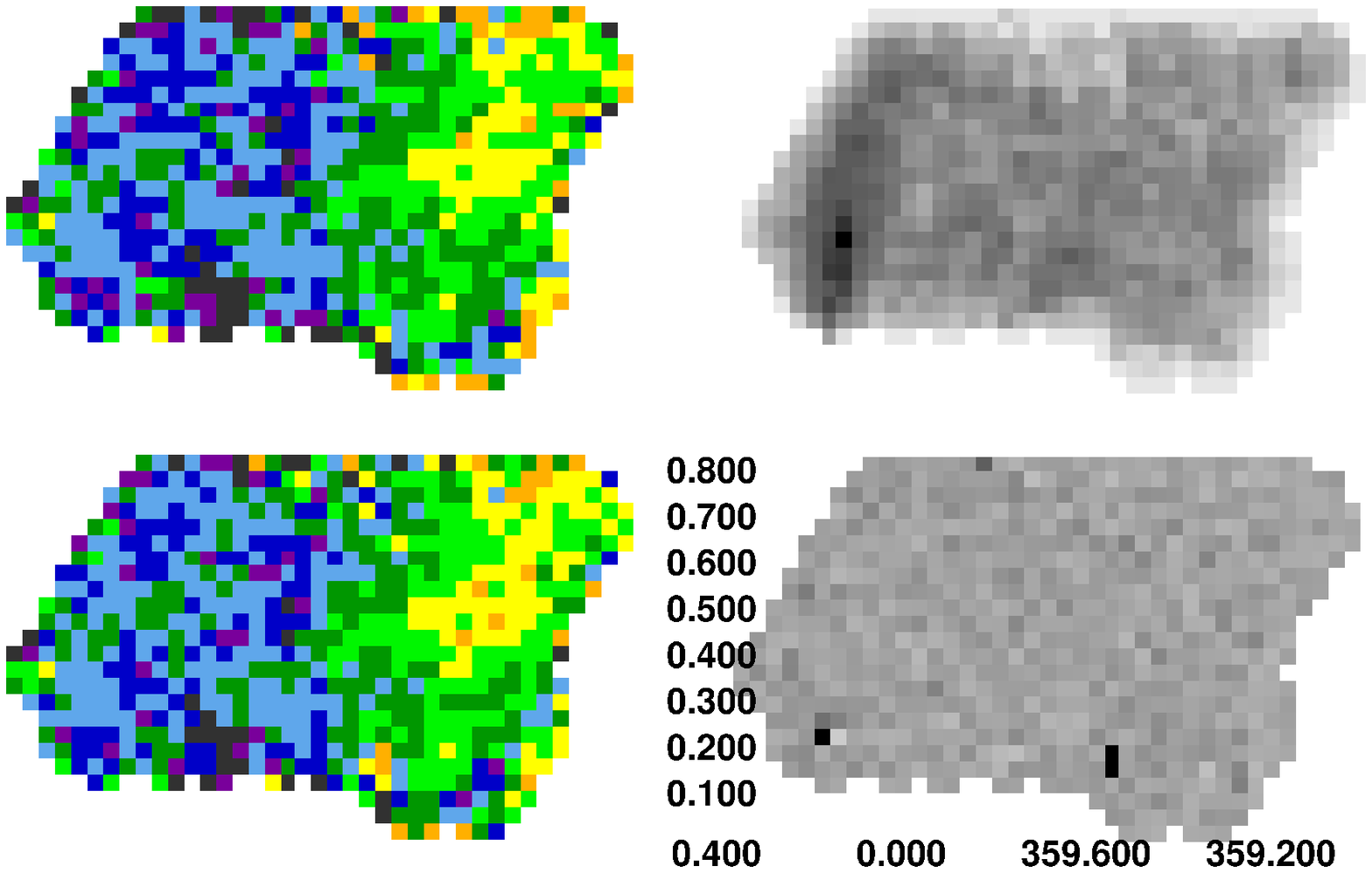}
\end{center}
\caption{\emph{Top left}: Smoothed map of \dt\ for the entire survey region with 125\arcsec$\times$125\arcsec\ tiles after using the histogram-fitting method.  The color-scale shows colors of light blue through purple representing negative \dt\ and colors from yellow through red representing positive \dt.  \emph{Top right}: Map of the error in \dt\ for the image shown in the top left.  The values range from 0\ddeg08 to 20\sdeg.  \emph{Bottom left}:  Smoothed map of \dt\ as in the top left panel when using the mean method. \emph{Bottom right}: Map of the difference significance between the histogram-fitting and mean methods with values ranging from --1$\sigma$\ to +1$\sigma$.  \label{padicomp}}
\end{figure}

Figure \ref{padicomp} compares images of \dt\ using the mean and histogram-fitting methods for a grid of 125\arcsec\ boxes.  An image showing the significance of differences between the two methods in each pixel.  The average difference between the two images is 0\ddeg002 and the standard deviation is about 0\ddeg2, so there is no significant difference between the two methods.

The image showing the difference in the two smoothing methods highlights two regions with greater than 1$\sigma$:  $(359\ddeg55,0\ddeg15)$ and $(0\ddeg15,0\ddeg2)$.  These differences demonstrate the biases of the methods.  In both locations, there is a small, polarized source with a different \dt\ from the large, polarized background.  Since the mean method favors bright sources and the histogram-fitting method favors large sources, these two regions are most likely to show a difference.  While this is an important caveat, Figure \ref{padicomp} shows that, in general, the two methods are in good agreement.

The errors on \dt\ differ according to the method used.  When averaging over 125-arsec tiles, both methods give errors within 50\% of each other, if errors $<3$\sdeg.  However, for errors greater than 3\sdeg, the errors found with the mean method are progressively less than those found by histogram fitting.  The maximal error found by the mean is about 20\sdeg, while the histogram-fitting method has a maximal error of about 200\sdeg.  This is expected, since the mean method assumes that angles are much smaller than 1 rad;  this is not always true for the errors in \dt.  Thus, the histogram fitting errors are a more accurate and more conservative estimate of the true errors in the mean value of \dt.
\end{appendix}

\end{document}